\documentclass[aps,floatfix,preprint,showpacs,preprintnumbers,nofootinbib,superscriptaddress,natbib]{revtex4}
\pdfoutput=1

\usepackage{graphicx,float}
\usepackage{epsfig}		
\usepackage{dcolumn}
\usepackage{bm}
\usepackage{subfigure}

\usepackage[all]{xy}
\usepackage{amsmath,upgreek}
\usepackage{amssymb}

\usepackage{pdfpages}
\usepackage{color}
\usepackage{graphicx,epstopdf}
\usepackage[colorlinks=true,
 linkcolor=red,
 urlcolor=purple,
 citecolor=blue,hyperindex]{hyperref}
\def\0{\mbox{\tiny $0$}}
\def\1{\mbox{\tiny $1$}}
\def\2{\mbox{\tiny $2$}}
\def\3{\mbox{\tiny $3$}}
\def\4{\mbox{\tiny $4$}}
\def\5{\mbox{\tiny $5$}}
\def\6{\mbox{\tiny $6$}}
\def\7{\mbox{\tiny $7$}}
\def\8{\mbox{\tiny $8$}}
\def\9{\mbox{\tiny $9$}}

\def\f14{\mbox{\tiny $\frac{1}{4}$}}

\def\bb#1{\mbox{\footnotesize $(#1)$}}

\begin{document}

\title{Quantum to classical transition in the Ho\v{r}ava-Lifshitz quantum cosmology}
\author{A. E. Bernardini}
\email{alexeb@ufscar.br}
\altaffiliation[On leave of absence from]{~Departamento de F\'{\i}sica, Universidade Federal de S\~ao Carlos, PO Box 676, 13565-905, S\~ao Carlos, SP, Brasil.}
\author{P. Leal}
\email{up201002687@fc.up.pt}
\altaffiliation[Also at~]{Centro de F\'isica do Porto, Rua do Campo Alegre 687, 4169-007, Porto, Portugal.} 
\affiliation{Departamento de F\'isica e Astronomia, Faculdade de Ci\^{e}ncias da
Universidade do Porto, Rua do Campo Alegre 687, 4169-007, Porto, Portugal.}
\author{O. Bertolami}
\email{orfeu.bertolami@fc.up.pt}
\altaffiliation[Also at~]{Centro de F\'isica do Porto, Rua do Campo Alegre 687, 4169-007, Porto, Portugal.} 
\affiliation{Departamento de F\'isica e Astronomia, Faculdade de Ci\^{e}ncias da
Universidade do Porto, Rua do Campo Alegre 687, 4169-007, Porto, Portugal.}
\date{\today}

\begin{abstract}
A quasi-Gaussian quantum superposition of Ho\v{r}ava-Lifshitz (HL) stationary states is built in order to describe the transition of the quantum cosmological problem to the related classical dynamics.
The obtained HL phase-space superposed Wigner function and its associated Wigner currents describe the conditions for the matching between classical and quantum phase-space trajectories.
The matching quantum superposition parameter is associated to the total energy of the classical trajectory which, at the same time, drives the engendered Wigner function to the classical stationary regime.
Through the analysis of the Wigner flows, the quantum fluctuations that distort the classical regime can be quantified as a measure of (non)classicality.
Finally, the modifications to the Wigner currents due to the inclusion of perturbative potentials are computed in the HL quantum cosmological context.
In particular, the inclusion of a cosmological constant provides complementary information that allows for connecting the age of the Universe with the overall stiff matter density profile.
\end{abstract}

\date{\today}
\maketitle

\section{Introduction}

Our ideas about space-time provided by General Relativity (GR) are fundamentally linked to the classical dynamics of objects and particles.
For arbitrarily small scales, say at Planck scale, space-time can be considerably different.
The transition between the ultimately quantum picture and the large scale properties of the Universe can depend on fundamental assumptions related to quantum cosmology and quantum gravity \cite{Kiefer07,Bojowald}.
Such issues concern the very nature of time itself and of quantum probabilities \cite{Page83,Vilenkin94,Ander12,Merali13,Genovese}, as well as the cosmological boundary features and the initial singularity problem, and the ensued conditions for the onset of inflation \cite{Linde84,Rubakov84,Vilen84,Bousso96,Linde98,Hartle83, Grishchuk}. 

Aiming to set up a theory of quantum gravity, the canonical Hamiltonian formulation of the quantum cosmology driven by the Wheeler-DeWitt (WdW) equation for a wave function of the Universe \cite{DeWitt67} provides interesting scenarios to examine quantum effects in cosmology.
However, the absence of a complete definition for the specific properties of quantum states resulting from this framework is in itself an obstacle for a detailed modeling of the primordial conditions of the Universe.
It follows that this set up does not provide much room for observational implications.

The canonical WdW framework itself gives rise to various conceptual questions. 
One of them concerns the classification of infinite dimensional deformed algebras related to such a constraint equation, a discussion particularly relevant in loop quantum gravity \cite{Rovelli}.
On more physical grounds, the questions about the nature of time \cite{Page83,Vilenkin94,Genovese,VilenBerto1,VilenBerto2} stems from the conditions from which GR yields a constraint equation \cite{Hartle83,DeWitt67}, the WdW equation, and how it provides a clear-cut prescription for the time evolution.

In fact, classical cosmology does provide a robust insight into the understanding of the cosmic inventory and does allow for extracting from the phenomenological data quite relevant information about the existence of a dark sector that dominates the cosmic budget.
Therefore, a procedure to fit the transition from quantum to classical descriptions of the cosmological framework is particularly relevant.
In such a context, our proposal in this work is to examine the classical cosmology arising from the Ho\v{r}ava-Lifshitz (HL) \cite{Hora09,Horava:2010zj,Blas:2009qj,Blas:2010hb} cosmological models \cite{Mukohyama:2010xz,Saridakis:2011pk,Zarro,Mukohyama:2009mz,Soti2009bx,Maeda:2010ke,Wang:2009rw,Wang:2009rw2}. Starting from quantum state solutions obtained from the HL WDW equation, an extension to the phase-space Wigner formalism for quantum mechanics (QM) is set up in order to provide a formulation to track the transition from quantum to classical descriptions.

As is well known, the derivation of testable predictions from quantum cosmology requires systematic approximations at semiclassical level and beyond, and most often, the minisuperspace approximation arises from symmetry arguments to reduce the number of degrees of freedom.
Quantum cosmology in the context of minisuperspace models allows for the construction and investigation of the behavior of the wave function of the Universe obtained from the WdW equation.
This framework deals with the relativistic space-time without a unique Hamiltonian to generate the dynamics. In any case, despite the fact that one has several choices for the definition of time and the ensued Hamiltonian evolution, these choices should lead to the same physics.

Since the considered symmetries are respected at classical level and that at quantum level they lead to a set of consistency conditions that restricts the possible choices of quantum cosmologies, in the HL cosmologies \cite{Hora09} to be considered in this work, we shall be able to quantify some of the above-mentioned issues and to create a platform for describing quantum to classical transition scenarios.

Essentially, one assumes that the cosmological modifications due to QM can be computed from an extended version of the Liouville equation which describes the dynamical evolution of a quasi-Gaussian Wigner function in the Weyl-Wigner framework of QM.
This framework depicts quite well the interplay between quantum and classical variables as it effectively takes place in constrained systems.

The key issue in the construction of the time-evolution of the quantum Wigner functions and of their associated probability currents is that it provides a quite illustrative picture of the quantum to classical transitions for the solutions of the WdW equation.
In this context, the non-trivial configuration of initial quasi-Gaussian quantum states, as Hamiltonian eigenstate superpositions of projectable gravity scenarios according to HL cosmologies \cite{Hora09}, can be investigated and their effects analytically quantified.
For some simplified HL cosmological dynamics, the quantum to classical transition can be characterized by quantifiers of quantum distortion and nonclassicality, computed from Wigner currents.

Our analysis uses as a starting point the projectable HL gravity without detailed balance in a minisuperspace quantum cosmological Friedmann-Lema\^{\i}tre-Robertson-Walker (FLRW) universe without matter\footnote{
This model exhibits the main features of the HL gravity, in a completely covariant approach \cite{Horava:2010zj}, which provides the detailed balance as a limiting hypothesis, being much simpler than non-projectable models \cite{Blas:2009qj,Blas:2010hb}.}.
In fact, several cosmological scenarios have been investigated in the context of HL gravity \cite{Mukohyama:2010xz,Saridakis:2011pk,Zarro}, where projectable versions are those where the lapse function depends uniquely on time.
This implies that the non-local classical Hamiltonian constraint of the GR must be integrated over spatial coordinates.
This leads to modifications to the Friedman equation through an additional term that behaves like dust \cite{Mukohyama:2009mz}, although suppressed by integration \cite{Soti2009bx}.
Concomitantly, higher spatial curvature terms give rise to new cosmological features \cite{Maeda:2010ke,Wang:2009rw,Wang:2009rw2} involving, for instance, bouncing and oscillating solutions \cite{Kiritsis:2009sh,Brandenberger:2009yt,Calcagni:2009ar}.

In what follows we shall consider the quantum cosmology of the HL quantum gravity in the minisuperspace approximation \cite{Zarro} and study its transition to classical cosmology. The onset of our study is the hyperbolic WdW equation \cite{Halliwell} and its solution for the HL gravity \cite{Zarro,Garattini:2009ns}, where the cosmological constant and matter components are included in the analysis \cite{Soti2010wn}.

It will be shown that in phase-space the quantum system is akin to the classical behavior, even though quantum properties are essential for a consistent description.
In such a context, it is relevant to notice that the transition from quantum to classical behavior is often described with the phase-space Wigner formalism by decoherence processes \cite{22,23,26}. 
Our analysis, however, provides results in terms of the Wigner currents and shows, through analytical expressions for the cosmological classical limit, an exact correspondence with the quantum cosmological description.
The existence of such a quantum to classical correspondence is supported by the coincidence of classical trajectories with sharp peaks of the Wigner function, the most likely quantum region arising from a quasi-Gaussian expression obtained from the superposition of HL quantum states.

The outline of this work is as follows.
Section II is concerned with the HL quantum cosmologies in the minisuperspace approximation as reported in the literature.
The WdW equation is re-obtained and a class of solutions which contain radiation, curvature and stiff matter contributions is worked out.
In Section III, an analytical expression for the Wigner function with quasi-Gaussian profile is obtained as a superposition of the HL solutions. 
It is pointed out that the parameters that drive the quantum superposition can be adjusted as to analytically fit the classical trajectories associated to the corresponding classical cosmology.
The dynamics of such a quantum to classical transition can be discussed in terms of the Wigner flow analysis (cf. Appendices I and II) presented in Section IV, where quantum distortions are exactly obtained from defined Wigner currents. In particular, modifications due to the extension of the formalism to the description of models with bounces are also discussed.
The inclusion of a cosmological constant density profile component is performed in Section V in order to modify the Wigner currents which, in this case, are perturbatively re-obtained.
A detailed analysis of the corresponding Wigner flow provides an expression for computing the age of the Universe in terms of the stiff matter contribution.
Our concluding remarks are drawn in Section VI and they emphasize the efficiency of the Wigner formalism in describing cosmological quantum to classical transitions.

\section{HL quantum cosmologies in the minisuperspace limit}

The setup for our framework is the Einstein-Hilbert action given by
\begin{equation}
\mathcal{S} = \frac{1}{16 \pi G}\int{d^{4}x\, \sqrt{-\mathtt{g}}\, \mathcal{R}},
\label{eqn00}
\end{equation}
where $\mathtt{g} = \det{(g_{\mu \nu})}$, $\mathcal{R} = R^{\mu \nu}\,g_{\mu \nu}$ is the scalar curvature, and $c = \hbar = 1$.
The most general form of a $SO(4)$-invariant metric in a $M = \mathbb{R} \times S^3$ topology \cite{BertolamiMourao}, in an homogeneous and isotropic space-time, is given by the line element of the Robertson-Walker (RW) metric,
\begin{equation}
ds^{2} = - \sigma^{2}\left[N(t)^{2}\,dt^{2} - a(t)^2\,\left(\frac{dr^2}{1- g_{_{C}} r^2} + r^2\,d\Omega^{2}\right) \right] ,
\label{eqn01}
\end{equation}
where $\sigma$ is a normalization constant, and $g_{_{C}} =0,\, +1$, and $-1$ denotes the curvature corresponding to $\mathbb{R}^3$, $S^3$ and $H^3$ hypersurfaces.
In this case, one has $\sqrt{-\mathtt{g}} = N(t) \, a(t)^3$, where the lapse function, $N(t)$, and the scale parameter, $a(t)$, are arbitrary non-vanishing functions of time, $t$, and $d\Omega^{2} = d\theta^2 + \sin{(\theta)}^2 d\phi^2$.

More generically, in terms of metric components, the three-dimensional quantities used to describe the GR in the Arnowitt-Deser-Misner (ADM) formalism \cite{Misner,ABOB01} are
$g_{ij}$, $\Pi_{ij} = \sqrt{-g}\left(\Gamma_{kl}^0 - g_{kl}\,\Gamma_{mn}^0\, g^{mn}\right)g^{ik}g^{jl}$,
$N = (-g^{00})^{-{1}/{2}}$, and the shift vector, $N_{i} = g_{0i}$, where one has the connection, $\Gamma_{ij}^{k}$, as an independent quantity, with {\em latin} indices running from $1$ to $3$.
Relevant to the discussion is the extrinsic curvature written as
\begin{equation}
K_{ij}=\frac{1}{2\sigma N}\left(-\frac{\partial g_{ij}}{\partial t}+\nabla_{i}N_{j}+\nabla_{j}N_{i}\right),
\label{eqn03}
\end{equation}
where $\nabla_{i}$ denotes the $3$-dimensional covariant derivative, and for the metric Eq.~\eqref{eqn01},
\begin{equation} 
K_{ij}=-\frac{1}{\sigma N} \frac{\dot{a}}{a} g_{ij}, \quad \mbox{with} \quad K = K^{ij}g_{ij} = -\frac{3}{\sigma N} \frac{\dot{a}}{a},
\label{eqn04}
\end{equation}
as $N_i = 0$ (and $g_{_C}= 1$).
In this case, the $3$-dim Ricci tensor components and the corresponding Ricci scalar are given, respectively, by $R_{ij}={2 g_{ij}}/{(\sigma^{2}a^{2})}$ and $R={6}/{\sigma^{2}a^{2}}$.

Finally, an interesting suggestion to achieve a renormalizable quantum gravity theory at high-energy is provided by the HL gravity, which at cosmological level is given by the action \cite{Soti2009gy,Soti2009bx}, 
\begin{equation}
\begin{split}
\mathcal{S}_{HL}&=\frac{M_{\mbox{\tiny Pl}}^{2}}{2} \int d^{3}x \,dt \,N\sqrt{g} \left.\bigg{\{}K_{ij}K^{ij}-\lambda K^{2} -g_{0}M_{\mbox{\tiny Pl}}^{2} -g_{1}R\right.\\ &\left. \qquad\qquad -g_{2}M_{\mbox{\tiny Pl}}^{-2}R^{2} -g_{3}M_{\mbox{\tiny Pl}}^{-2}R_{ij}R^{ij}
 -g_{4}M_{\mbox{\tiny Pl}}^{-4}R^{3}-g_{5}M_{\mbox{\tiny Pl}}^{-4}R\left(R^{i}_{\;\,j}R^{j}_{\;\,i}\right) 
 \right.\\ &\left. \qquad\qquad\qquad\qquad
 -g_{6}M_{\mbox{\tiny Pl}}^{-4}R^{i}_{\;\,j}R^{j}_{\;\,k}R^{k}_{\;\,i} - g_{7}M_{\mbox{\tiny Pl}}^{-4}R\nabla^{2}R -g_{8}M_{\mbox{\tiny Pl}}^{-4}\nabla_{i}R_{jk}\nabla^{i}R^{jk}\right.\bigg{\}},
\end{split}
\label{eqn06}
\end{equation}
where the balance of the curvature components is described by dimensionless coupling constants, $g_{i}$, with $i = 0,\dots,9$, and $M_{\mbox{\tiny Pl}}$ denotes the Planck mass. 
Here it is worth to mention that the HL gravity is a framework in which the ultraviolet (UV) completion problem of GR \cite{Hora09} is circumvented by turning the gravity into a power-countable renormalizable theory at the UV fixed point.
This is achieved by giving up the Lorentz symmetry at high-energies \cite{Hora09,Visser:2009fg}, assuming that GR is recovered at an infra-red (IR) fixed point scenario of the HL gravity at low-energy scales.
Such a Lorentz symmetry breaking is related to an anisotropic scaling of space and time, $r\rightarrow b r$ and $t \rightarrow b^{z} t$, with $b$ being a scale parameter and $z\neq1$. 
 
A sequence of assumptions carried out in Ref.~\cite{Zarro} did reduce the number of degrees of freedom of the action Eq.~\eqref{eqn06}.
Firstly, one notices that the GR is recovered by setting $g_{1}=-1$, through the rescaling of the time coordinate, and by taking the limit of $\lambda\rightarrow 1$, which recovers the full diffeomorphism invariance.
However, $\lambda$ must be a running constant, and there is no reason or symmetry that constrains it to the GR limit\footnote{Even if the phenomenology suggests that $\lambda$ is quite close to unity \cite{Soti2010wn}.}.
As for the cosmological constant, $\Lambda$, it can be written in Planck units as $\Lambda = g_{0} M_{\mbox{\tiny Pl}}^{2}/2$. 
Finally, isotropic and homogeneous conditions over $g_{ij}$ impose constraints directly into the equations of motion, or through the substitution of the RW metric into the Lagrangian density\footnote{In Ref.~\cite{Zarro} the authors point to that such restrictions cannot be done over the Lagrangian unless one properly solves the arising constraints. Otherwise, the introduction of the RW metric does not lead, in general, to the same results obtained from the equations of motion \cite{Bertolami:1990je}.}. 
Since the RW metric, Eq.~(\ref{eqn01}), introduces an anisotropy between space and time, it does not affect the homogeneity. Then, it can be substituted into Eq.~(\ref{eqn06}) and the integration over $d^{3}x \sqrt{\det{g_{ij}}}$ gives $2\pi^{2}$ so that the HL minisuperspace action \cite{Zarro} is re-written as
\begin{equation}
\begin{split}
S_{HL}&=\frac{M_{\mbox{\tiny Pl}}^{2}\times2\pi^{2}\times 3(3\lambda -1)\sigma^{2}}{2} \int dt N \left\{\frac{-\dot{a}^{2}a}{N^{2}} +\frac{6a}{3(3\lambda-1)} -\frac{2\Lambda\sigma^{2}a^{3}}{3(3\lambda-1)}-\right. \\
&\left. -M_{\mbox{\tiny Pl}}^{-2}\times\frac{12}{3(3\lambda-1)\sigma^{2}a}\times (3g_{2}+g_{3}) -M_{\mbox{\tiny Pl}}^{-4}\times\frac{24}{3(3\lambda-1)\sigma^{4}a^{3}}\times (9g_{4}+3g_{5}+g_{6})\right\}.
\end{split}
\label{eqn07}
\end{equation}
By redefining the dimensionless constants \cite{Maeda:2010ke}
\begin{eqnarray}
 g_{_{C}} &=& \frac{2}{3\lambda-1},\\
 g_{_{\Lambda}} &=& \frac{\Lambda M_{\mbox{\tiny Pl}}^{-2}}{9\pi^{2}(3\lambda-1)^{2}},\\
 g_{_{R}} &=& 24\pi^{2} (3g_{2}+g_{3}), \\
 g_{_{S}} &=& 288\pi^{4}(3\lambda -1) (9g_{4}+3g_{5}+g_{6}),
\label{eqn09} 
\end{eqnarray}
and choosing units simplified by the constraint $\sigma^{2}\times6\pi^{2}\times(3\lambda-1)M_{\mbox{\tiny Pl}}^{2}=1$, the minisuperspace action finally reads
\begin{equation}
S_{HL}=\frac{1}{2}\int dt \left(\frac{N}{a}\right)\left[-\left(\frac{a}{N}\dot{a}\right)^{2} +g_{_{C}}a^{2} -g_{_{\Lambda}}a^{4}-g_{_{R}}-\frac{g_{_{S}}}{a^{2}} \right],
\label{eqn10}
\end{equation}
from which one identifies the canonical conjugate momentum associated to $a$ as given by
\begin{equation}
\Pi_{a}=\frac{\partial \mathcal{L}}{\partial \dot{a}}= - \frac{a}{N}\dot{a},
\label{eqn11}
\end{equation}
such that the HL minisuperspace Hamiltonian density becomes \cite{Soti2009bx,Zarro}
\begin{equation}
 H=\Pi_{a}\dot{a}-\mathcal{L}=\frac{1}{2}\frac{N}{a}\left(-\Pi_{a}^{2}-g_{_{C}}a^{2} +g_{_{\Lambda}}a^{4}+g_{_{R}}+\frac{g_{_{S}}}{a^{2}}\right).
\label{eqn12}
\end{equation}

A discussion of the quantum mechanical problem resulting from the above Hamiltonian in the context of the WdW framework is presented in Refs. \cite{Soti2009bx,Soti2010wn,Hora09,Visser:2009fg,Zarro}. One notices that $g_{_{C}} > 0$ stands for the curvature coupling constant and the sign of $g_{_{\Lambda}}$ follows the sign of the cosmological constant. In addition, both of them, $g_{_{C}}$ and $g_{_{\Lambda}}$, are related to each other through a common degree of freedom, $\lambda$, such that the limit for the minisuperspace GR model is recovered by setting $\lambda = 1$.
The sign of the coupling constants $g_{_{S}}$ and $g_{_{R}}$, associated to stiff matter and radiation like contributions, respectively \cite{Soti2009bx}, does not affect the stability of the HL gravity \cite{Maeda:2010ke,Hora09}.

By following the canonical quantization strategy \cite{DeWitt67,Hartle83}, the canonical conjugate momentum is promoted to an operator \cite{Hartle83}, 
$$\Pi_{a}\mapsto -i \frac{d}{d a} \quad\mbox{such that} \quad\Pi_{a}^{2}=-\frac{1}{a^{q}}\frac{d}{d a}\left(a^{q}\frac{d}{d a}\right),$$
where the choice of $q$ does not affect the semiclassical analysis \cite{KolbTurner:1989}.
The classical minisuperspace Hamiltonian is thus promoted to an operator which acts on the wave function of the Universe $\tilde{\psi}(a)$ such that the final form of the WdW equation (for $q = 0$) is then given by
\begin{equation}\label{eqn14}
\frac{1}{2}\left(\frac{d^{2}}{d a^{2}}-g_{_{C}}a^{2} +g_{_{\Lambda}}a^{4}+g_{_{R}}+\frac{g_{_{S}}}{a^{2}}\right)\tilde{\psi}(a)=0,
\end{equation}
from which one identifies the quantum potential given by 
\begin{equation}
V(a)=\frac{1}{2}\left(g_{_{C}}a^{2}-g_{_{\Lambda}}a^{4}-g_{_{R}}-\frac{g_{_{S}}}{a^{2}}\right).
\label{eqn15}
\end{equation}
This is an one-dimensional Schr\"{o}dinger-like equation constrained by a vanishing eigenvalue, $E = 0$, for which a complete analysis of its eigenvalues has already been performed in Refs.~\cite{Zarro,Maeda:2010ke}.
However, the constraint, $E = 0$, in a certain sense, blurs the understanding of the {\em meaning of time} in the above procedure.
Given that the classical time evolution is set by the Friedmann equation, any discussion of quantum to classical transition should take into account the conditions for the matching between the quantum and the classical frameworks.

In quantum cosmology, the notion of time is established following different premises since, in GR, time is observer-dependent and not absolute, and time translations are not generated by an observable, such as the energy, but by an expression which, on physical states, is constrained to vanish.
Time translations are part of the transformations allowed by the algebra of the fundamental space-time symmetries \cite{Burg,Rovelli,Bojowald}.
Obtaining a consistent time definition stems from the evolution that a system experiences after imposing some smearing conditions into the canonical Hamiltonian formalism \cite{117,118,119}.
The simplest way to establish a time evolution, for specific matter/field contributions, is through the so called de-parameterization procedure: one simply picks the variable that classically depends monotonically on time and interpret it as time itself \cite{Burg,tempo,Bojowald}.
On more general grounds, the absence of a covariant treatment leads to different quantum theories and one cannot find a unitary transformation to consistently identify how the observables vary \cite{120,121,122,BertolamiTime1,BertolamiTime2,BertolamiTime3}.

Given these difficulties, it is sensible to expect that the choice of the {\em time parameter} should not affect the physical results.
The point in this manuscript, which shall be postponed to the discussion of the quantum to classical transitions at the end of Section III, is that the identification of a parametric canonical variable, $\tau_{}$, with the quantum mechanical time, implicitly identified by a Hamiltonian correspondence with $i \,\partial/\partial \tau_{}$, should be a suitable map for the {\em classical} results so to establish a natural bridge between quantum to classical cosmological descriptions.
The association of the time parameter with the radiation energy has a clearly classical appeal since, classically, the correspondence between time and inverse square temperature of the cosmic background radiation is consistent with the phenomenological analysis that accounts for the cosmic energy density inventory.
Although lacking a covariant framework, the association between time and temperature is consistent with suggestions for the origin of time asymmetry according to which the arrow of time does not reverse at an eventual contraction of the Universe \cite{Laflamme,Laflamme2}.
A consistent definition for such an extra degree of freedom, $\tau_{}$, should constrain the properties of the Hamiltonian operator. 
Herein, the notion of an extra degree of freedom associated to an environment set by radiation, with an associated unitary operator, $H_{\nu} \equiv i \partial/\partial \tau_{}$, for the coordinate $\tau_{}$, allows one to rewrite the wave function as $\tilde{\psi}(a) \equiv \tilde{\psi}(a,\tau_{})$ as to have
\begin{equation}
H_{\nu} \tilde{\psi}(a,\tau_{}) = E \tilde{\psi}(a,\tau_{}),\quad \Rightarrow \quad \tilde{\psi}(a,\tau_{}) = \psi(a)\,\exp(-i\,E\,\tau_{}),
\label{eqn13}
\end{equation}
with $E$ and $\tau_{}$ in Planck units (cf. after Eqs.~(\ref{eqn07})-(\ref{eqn09})).
It is important to point out that, in our approach, the correspondence between time and radiation energy provides the elements for composing a large set of energy eigenstate quantum superpositions that exhibit a Schr\"odinger-like time evolution.
In fact, the identification of the time variable in quantum cosmology is associated to the transition from quantum to classical dynamics \cite{Habib02} where several competing frameworks have been considered, either for pure states or for open systems.
Our assumption resembles operationally the effects of unimodular quantum cosmology \cite{Unruh} where time arises 
from a secondary constraint which, once integrated, leads to an expression similar to that of Eq.~(\ref{eqn13}).
In that case, the associated energy, $E$, is actually the cosmological constant, and the time parameter is extrinsically identified with the Hamiltonian operator. As in our approach, the time is not an internal dynamical unconstrained phase-space variable.
Otherwise, besides the usual WdW Hamiltonian constraint written in the form of Eq.~(\ref{eqn13}), in our approach there is no additional secondary constraint as considered in the unimodular framework.
Such an interpretation for $\tau_{}$ is not different from the inclusion of some canonical coordinate related to additional matter/field/radiation components into the action, with the attribute of a time variable \cite{tempo,Pedram,Blyth}.

In fact, the parameter $g_{_{R}}$ can be rewritten as $g_{_{R}} = g_{\gamma} + g_{\nu} = g_{\gamma} + 2E$, in order to give a physical meaning to the eigenvalue $E$, where the parameters $g_{\gamma}$ and $g_{\nu}$ are, for instance, identified with photon and with massless neutrino contributions, respectively.

By introducing the elements of the above discussion and setting $a = g_{_{C}}^{-{1}/{4}} x$, one obtains the re-parameterized equation
\begin{equation}\label{eqn16}
\left\{\frac{d^{2}}{d x^{2}} - x^{2} - \frac{4 \alpha^2 -1}{4 x^{2}} + 2 (\alpha+ 2n + 1) + \ell x^{4} \right\} \psi^{\alpha}_n(a\bb{x})=0,
\end{equation}
with $\ell = g_{_{C}}^{-3/2}\,g_{_{\Lambda}}$, and where one identifies the parameters $\alpha$ and $n$ related to $E \to E_n =g_{_{C}}^{{1}/{2}}\left(2n + 1\right)$, and to the coupling constants by $g_{_{R}} = 2\alpha\,g_{_{C}}^{{1}/{2}}$ and $g_{_{S}} = -(4\alpha^{2}-1)/4$, for the wave function given by
\begin{equation}
\psi^{\alpha}_n(a\bb{x}) = {g_{_{C}}^{\frac{1}{8}}} \varphi^{\alpha}_n(x).
\label{eqn18}
\end{equation}

An exact solution of the above quantum mechanical problem can be obtained for $g_{_{\Lambda}} = 0$.
The solution is also valid for $0 \lesssim a \ll 1$ for $g_{_{\Lambda}} \neq 0$ \cite{Zarro}, however this case demands for a perturbative treatment.
For $g_{_{\Lambda}} = 0$, one has
\begin{equation}
\varphi^{\alpha}_n(x) = 2^{{1}/{2}}\,\Theta(x) \, N_n(\alpha)\, x^{\alpha + \frac{1}{2}}\,\exp(-x^2/2)\,L^{\alpha}_n(x^2),
\label{eqn19}
\end{equation}
where $L^{\alpha}_n$ are the {\em associated Laguerre polynomials}, $\Theta(x)$ is the {\em step-unity function} that constrains the result to $x > 0$, and $N_n(\alpha)$ is the normalization constant given by
\begin{equation}
N_n(\alpha) = \left(\frac{n!}{\Gamma(n+\alpha+1)}\right)^{{1}/{2}}, 
\label{eqn20}
\end{equation}
where $\Gamma(s)$ is the {\em gamma function}.
\begin{figure}[h!]
\includegraphics[scale=0.8]{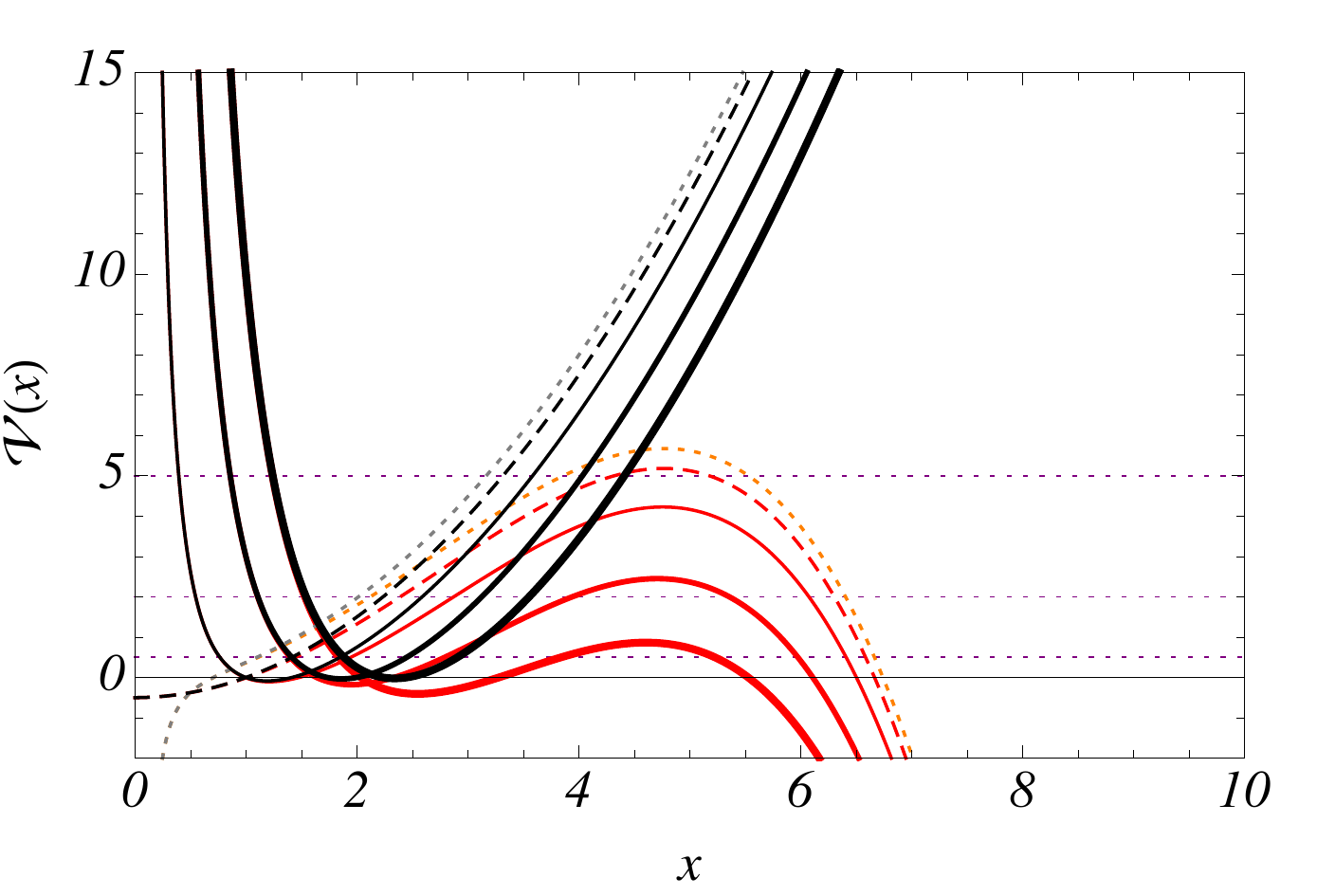}
\renewcommand{\baselinestretch}{.85}
\caption{
Re-parameterized potentials ${g_{_{C}}^{-{1}/{2}}}\,{V(a(x))} \equiv\mathcal{V}(x)$ as function of the cosmological parameter $x = g_{_{C}}^{{1}/{4}} a$ for the cosmological constant associated parameter $\ell = 0$ (black lines) and $\ell = 0.022$ (red lines).
The behavior of the red line plots do not qualitatively change for $\ell \neq 0$.
The plots are for $\alpha = 0$ (dotted lines), $1/2$ (dashed lines), $3/2$ (thinest solid lines), $7/2$ and $11/2$ (thickest solid lines). For $\alpha = 0$ one has $g_{_{S}} < 0$ and $g_{_{R}} = 0$ which leads to an undesirable singularity, and for $\alpha = 1/2$
one has no contribution from stiff matter, $g_{_{S}} = 0$, which corresponds to the one-dimensional harmonic oscillator limit in the Schr\"odinger equation.
Horizontal dashed lines refer to the classical energy correspondence for $E = 1/2,\, 2$ and $5$.}
\label{Figura001}
\end{figure}
The quantum mechanical potential, as depicted in Fig.~\ref{Figura001}, includes the contribution from the $g_{_{\Lambda}} \neq 0$ term, and it is now written as
\begin{equation}
{g_{_{C}}^{-{1}/{2}}}{V(a(x))} \equiv \mathcal{V}(x) = \frac{1}{2} \left[x^{2}+ \frac{4 \alpha^2 -1}{4 x^{2}} - 2 \alpha - \ell x^4\right].
\label{eqn17}
\end{equation}
Solution Eq.~(\ref{eqn19}) can be investigated in the framework of phase-space Wigner formalism in order to provide the elements that can be compared with the classical dynamics driven by $V(a) = g_{_{C}}^{{1}/{2}}\, \mathcal{V}(x)$.

\section{Wigner function for a quasi-Gaussian superposition and the classical limit}

The Wigner function and the Weyl transform establish an alternative framework to connect quantum observables to expectation values as it provides a suitable insight into the quantum behavior and its classical limit.
For a generic quantum state identified by the wave function, $f_{\alpha}(a)$, the Wigner function can be read as the Fourier transform of the off-diagonal terms of the associated density matrix, $f^*_{\alpha}(a+y_a)\,f_{\alpha}(a-y_a)$, which is given by the so called Weyl transform,
\begin{equation}
W^{\alpha}(a,\,\Pi_a) = \frac{1}{\pi}\int_{-\infty}^{+\infty}dy_a\, \exp(2i\,\Pi_a\,y_a) \,
f^*_{\alpha}(a+y_a)\,f_{\alpha}(a-y_a).\,
\label{eqn21}
\end{equation}
This is interpreted as a real-valued quasi-probability distribution, as $W^{\alpha}(a,\,\Pi_a)$ can, in principle, assume local negative values. This formulation has the operational advantage of exhibiting all the information content of the state vector in the phase-space, where operators, $\hat{a}$ and $\hat{\Pi}_a$, have been converted into $c$-numbers. 

A re-dimensionalized form of Eq.~(\ref{eqn21}) is written as
\begin{equation}
W^{\alpha}(a,\,\Pi_a) \equiv W^{\alpha}(x,\,p) = \frac{1}{\pi}\int_{-\infty}^{+\infty}dy\, \exp(2i\,p\,y) \,
\mathrm{F}^*_{\alpha}(x+y)\,\mathrm{F}_{\alpha}(x-y),
\label{eqn21B}
\end{equation}
where one has identified: $a = g_{_{C}}^{-{1}/{4}} x$, $y_a = g_{_{C}}^{-{1}/{4}} y$, $\Pi_a = g_{_{C}}^{{1}/{4}} p$, and $f_{\alpha}(a(x)) =g_{_{C}}^{1/8}\mathrm{F}_{\alpha}(x)$, such that
\begin{equation}
\int_{-\infty}^{+\infty}\hspace{-.3 cm}da\int_{-\infty}^{+\infty}\hspace{-.3 cm}d\Pi_a\, W^{\alpha}(a,\,\Pi_a) = \int_{-\infty}^{+\infty}\hspace{-.3 cm}dx\int_{-\infty}^{+\infty}\hspace{-.3 cm}dp\,W^{\alpha}(x,\,p) = 1.
\end{equation}

Generically speaking, to connect quantum operators to averaged observables, the trace of the product of two operators, $\hat{O}_1$ and $\hat{O}_2$, is given by the phase-space integral of the product of their Weyl transform \cite{Wigner,Case},
\begin{equation}
Tr_{\{x,p\}}\left[\hat{O}_1\hat{O}_2\right] = 
\int \hspace{-.15cm}\int \hspace{-.15cm} {dx\,dp} \,O^W_1(x, p)\,O^W_2(x, p),
\end{equation}
which, in terms of the properties of the density matrix operator, $\hat{\rho}$,
gives
\begin{equation}
Tr_{\{x,p\}}\left[\hat{\rho}\hat{O}\right] = \langle O \rangle = 
\int \hspace{-.15cm}\int \hspace{-.15cm} {dx\,dp}\,W(x, p)\,{O^W}(x, p),
\end{equation}
which establishes the natural generalization of the Wigner function, from pure states to statistical mixtures, with the purity $Tr[\hat{\rho}^2]$ given by
\begin{equation}
Tr_{\{x,p\}}[\hat{\rho}^2] = 2\pi\int \hspace{-.15cm}\int \hspace{-.15cm} {dx\,dp}\,W(x, p)^2,
\label{pureza}
\end{equation}
where the factor $2\pi$ is introduced in order to satisfy the constraints: $Tr[\hat{\rho}^2] = Tr[\hat{\rho}] = 1$, for pure states.

Getting back to the HL solutions with the assumed decoupling between space-like coordinate, $a$, and time, $\tau_{}$, as set by Eq.~\eqref{eqn13}, a time dependent version of $\mathrm{F}_{\alpha}(x)\sim \mathrm{F}_{\alpha}(x,\,\tau_{})$ is given by a superposition of the eigenstates from Eq.~(\ref{eqn19}) as
\begin{equation}
\label{supp}
\mathrm{F}_{\alpha}(x,\,\tau_{}) =\mathcal{N}_{_F} \sum_{n=0}^{\infty}c^{\alpha}_n(\tau_{})\, \varphi^{\alpha}_n(x),
\end{equation}
where $\mathcal{N}_{_F}$ is a normalization constant.
Once identifying the parameter $g^{{1}/{2}}_{_{C}}$ with an energy quantity, $\omega/2$, given in Planck units, $T_{\mbox{\tiny Pl}}^{-1}$, where $T_{\mbox{\tiny Pl}}$ is the Planck time, as to have $E_n = \omega\left(n + 1/2\right)$, one can construct a quasi-Gaussian superposition if $c^{\alpha}_n \equiv u^n\,N^{-1}_n(\alpha)\,\exp(- (i/2) \,\omega \tau_{})$, with $u\equiv u(\beta,\,\tau_{}) = \exp(-\beta + i\,\omega \,\tau_{})$, where $\omega \tau_{}$ is a dimensionless quantity, and $\beta$ is an arbitrary weight parameter which constrains the expansion coefficients to $|u| < 1$.
By substituting $\varphi^{\alpha}_n(x)$ into the above superposition, one obtains the complex function
\begin{eqnarray}
\mathrm{F}_{\alpha}(x,\,\tau_{}) 
&=& \mathcal{N}_{_F}\,\Theta(x)\, x^{\alpha + \frac{1}{2}}\,\exp(-x^2/2)\,\sum_{n=0}^{\infty}u^n\,L^{\alpha}_n(x^2)\nonumber\\
&=& \mathcal{N}_{_F}\,\Theta(x)\, x^{\alpha + \frac{1}{2}}\,(1-u)^{-(1+\alpha)}\exp\left[-\frac{1}{2}\left(\frac{1+u}{1-u}\right)x^2\right]
\label{eqn22}
\end{eqnarray}
with 
$$\mathcal{N}_{_F} =\left[\frac{(1- e^{-2\beta})^{1+\alpha}}{2\Gamma(1+\alpha)}\right]^{{1}/{2}}.$$
Then the normalized probability distribution associated to $\mathrm{F}_{\alpha}$ is explicitly written in terms of $\tau_{}$ and $\beta$ as
\begin{eqnarray}
|\mathrm{F}_{\alpha}(x,\,\tau_{})|^2 
&=& \frac{2\,\mu _{(\beta,\tau_{})}^{1+\alpha}}{\Gamma(1+\alpha)}\,\Theta(x)\,x^{1+2\alpha}\, \exp(-\mu _{(\beta,\tau_{})} x^2),
\label{eqnstart}\end{eqnarray}
with
\begin{equation}
\mu _{(\beta,\tau_{})}= \frac{\sinh(\beta)}{\cosh(\beta)-\cos(\omega \tau_{})}.
\label{eqn23}
\end{equation}
Such an auxiliary function, $\mu _{(\beta,\tau_{})}$, helps one to define a comoving coordinate, $\tilde{x}_{(\beta,\tau_{})}$ implicitly given by
\begin{equation}
\tilde{x}^2_{(\beta,\tau_{})} = \frac{\alpha+1}{\langle \mathrm{F}_{\alpha}(x,\,\tau_{})\vert x^2 \vert \mathrm{F}_{\alpha}(x,\,\tau_{}) \rangle}\,x^2 = \mu _{(\beta,\tau_{})} x^2, 
\label{eqn24}
\end{equation}
which defines the comoving profile of an infinitesimal element of the quasi-Gaussian probability, $dx\,|\mathrm{F}_{\alpha}(x,\,\tau_{})|^2 = d\tilde{x}\,|\mathrm{F}_{\alpha}(\tilde{x})|^2$, with
\begin{equation}
|\mathrm{F}_{\alpha}(\tilde{x})| = \frac{2}{\Gamma(1+\alpha)}\,\Theta(\tilde{x})\,\tilde{x}^{1+2\alpha}\, \exp(-\tilde{x}^2),\end{equation}
whose behavior is depicted in the first plot of Fig.~\ref{Figura002} for several values of $\alpha$.

From the above expression, one sees that it is easy to mathematically compose a Gaussian state from a superposition of $\mathrm{F}_{\alpha}$ functions in $\alpha$, which, however, might bring some unphysical features since different $\alpha$ values mix different stiff matter profiles. Nevertheless, the relevant point is that the obtained quasi-Gaussian profile reproduces the classical trajectories in the phase-space.

Through the calculation of the Wigner function, one can get $W_{\alpha}(x,p)$ for (semi)integer values of $\alpha$.
The less interesting case, of course, corresponds to $\alpha =1/2$ for which one has $g_{_{S}} = 0$ and the quantum mechanical potential is reduced to the harmonic oscillator one.
For $\alpha = 0$ one has $g_{_{S}} > 0$, with $g_\gamma = 0$.
Solutions for this case (for which the quantum mechanical potential is exhibited by Fig.~\ref{Figura001}) are not physically meaningful given the presence of a singularity at the origin, which compromises the evolution of the quantum eigenstates in terms of $\tau_{}$ to an asymptotic squeezed state around $a = 0$.

By substituting the expression from Eq.~(\ref{eqn22}) into Eq.~(\ref{eqn21B}) one obtains
\begin{eqnarray}
\label{eqn27}
W^{\alpha}(x,\,p;\,\tau_{}) &=& \frac{2\,\mu _{(\beta,\tau_{})}^{1+\alpha}}{\pi\,\Gamma(1+\alpha)}\,\int_{-\infty}^{+\infty}dy\,\Theta(x+y)\Theta(x-y)\,(x^2-y^2)^{\frac{1}{2}+\alpha}\nonumber\\
&&\qquad\qquad\qquad \exp\left(-\mu _{(\beta,\tau_{})} (x^2+y^2)\right)\,\exp\left(2\,i\,y(p + \tilde{\mu }_{(\beta,\tau_{})}\, x)\right)\\
&=& \frac{2\,\mu _{(\beta,\tau_{})}^{1+\alpha}}{\pi\,\Gamma(1+\alpha)}\,x^{2(1+\alpha)}
\exp\left(-\mu _{(\beta,\tau_{})} x^2\right)\,\nonumber\\
&&\qquad \int_{-1}^{+1}ds\, (1-s^2)^{\frac{1}{2}+\alpha}\, \exp\left(-\mu _{(\beta,\tau_{})}\,x^2\,s^2\right)\,\exp\left(2\,i\,x\,s (p + \tilde{\mu }_{(\beta,\tau)}\, x)\right),\nonumber
\end{eqnarray}
where the dependence on $\tau_{}$ has been included through $\mu _{(\beta,\tau_{})}$ (cf. Eq.~\eqref{eqn23}) and
\begin{equation}
\tilde{\mu }_{(\beta,\tau_{})} = -\frac{\sin(\omega \tau_{})}{\cosh(\beta)-\cos(\omega \tau_{})}.
\label{tilde}
\end{equation}
In Eq.~(\ref{eqn27}), the integrating variable has been simplified to $y\sim x \, s$, which is helpful for verifying the normalization conditions for $W^{\alpha}(x,\,p;\,\tau_{})$ (cf. the calculations performed in the Appendix I).
By observing the function parity over the symmetric limits, the integration can be further simplified
by introducing the power series expansion,
\begin{equation}
\cos(2z) = \sum_{k=0}^{\infty}(-1)^k\,\frac{2^{2k}}{(2k)!}z^{2k}, 
\end{equation}
and using
\begin{eqnarray}
\lefteqn{2\int_{0}^{+1}ds\, (1-s^2)^{\frac{1}{2}+\alpha}\, s^{2k}\,\exp\left(-\mu _{(\beta,\tau_{})}\,x^2\,s^2\right) =}\nonumber\\
&&\qquad\qquad
\Gamma(3/2+\alpha)\Gamma(1/2+k)\,\, _1\mathcal{F}_1(1/2+k,2+\alpha+k,-\mu _{(\beta,\tau_{})}\,x^2\,s^2),
\end{eqnarray}
where $_1\mathcal{F}_1$ is the {\em confluent hypergeometric function of the first kind}. 
One thus obtains the final form of the Wigner function as
\begin{eqnarray}
\label{finalform}
W^{\alpha}(x,\,p;\,\tau_{}) &=& \frac{2}{\sqrt{\pi}}\frac{\Gamma(3/2+\alpha)}{\Gamma(1+\alpha)}
(\mu _{(\beta,\tau_{})}x^2)^{1+\alpha}
\exp\left(-\mu _{(\beta,\tau_{})} \,x^2\right)\\
&&\qquad
\sum_{k=0}^{\infty}\frac{(-1)^k\,(p\,x + \tilde{\mu }_{(\beta,\tau)}\, x^{2})^{2k}}{\Gamma(1+k)\,\Gamma(2+k+\alpha)}\,\,_1\mathcal{F}_1(1/2+k,2+\alpha+k,-\mu _{(\beta,\tau_{})}\,x^2\,s^2),\nonumber
\end{eqnarray}
which is not very helpful given that it involves an infinite sum over $k$.

However, for semi-integer values of $\alpha$, written as $\alpha = 1/2 + \upsilon$, with $\upsilon = 0,\,1,\,2,\, \dots$, one has the finite sum
\begin{equation}
(1-s^2)^{\frac{1}{2}+\alpha} = (1-s^2)^{1 + \upsilon} = \sum_{k=0}^{1 + \upsilon}
(-1)^k\,\frac{s^{2k}(1+\upsilon)!}{k!(1+\upsilon-k)!} = \sum_{k=0}^{1 + \upsilon}
\frac{\Gamma(3/2+\alpha)}{\Gamma(3/2+\alpha-k)\Gamma(1+k)},
\end{equation}
which, after substitution into Eq.~(\ref{eqn27}), leads to an expression that is easier to handle,
\begin{eqnarray}
\label{finalform2}
W^{\alpha}(x,\,p;\,\tau_{})
&=& \frac{4}{\pi}\frac{\Gamma(3/2+\alpha)}{\Gamma(1+\alpha)}
(\mu \,x^2)^{1+\alpha}
\exp\left(-\mu \,x^2\right)\sum_{k=0}^{\frac{1}{2}+\alpha}\frac{(-1)^k}{\Gamma(1+k)\,\Gamma(3/2+k+\alpha)}\,
\,\nonumber\\
&&\qquad \int_{0}^{+1}ds\, s^{2k}\,\exp\left(-\mu \,x^2\,s^2\right)\,\cos\left(2\,x\,s (p + \tilde{\mu }\, x)\right)\nonumber\\
&=& \frac{1}{\sqrt{\pi}}\frac{\Gamma(3/2+\alpha)}{\Gamma(1+\alpha)}
(\mu \,x^2)^{1+\alpha}
\exp\left(-\mu \,x^2\right)
\,\sum_{k=0}^{\frac{1}{2}+\alpha}\frac{x^{-(1+2k)}}{\Gamma(1+k)\,\Gamma(3/2+k+\alpha)}\nonumber\\
&&\qquad \,
\frac{d^k}{d\mu ^k}\left[\mu ^{-{1}/{2}} \exp\left[-\frac{(p+\tilde{\mu }x)^{2}}{\mu }\right]
\left(\mbox{Erf}[\zeta(\mu ,\tilde{\mu })] + h.c.\right)
\right],
\end{eqnarray}
with $\zeta(\mu ,\tilde{\mu }) = \mu ^{{1}/{2}}(x + i\,\mu ^{-1} (p+\tilde{\mu }x))$, and where the subindex $_{(\beta,\tau_{})}$ has been suppressed \footnote{To go further, given that the {\em Error function}, Erf$[\dots]$, has analytically well-defined derivatives written in terms of {\em Hermite polynomials}, $\mathcal{H}_{k-1}(z)$,
\begin{equation}
 \frac{\sqrt{\pi}}{2} \frac{d^k}{dz^k}\mbox{Erf}(z) = (-1)^{k-1}\, \mathcal{H}_{k-1}(z)\,\exp(-z^{2}),
\end{equation}
one can write
\begin{eqnarray}
\mbox{Erf}\left[\mu ^{{1}/{2}}\left(x + i\frac{p+\tilde{\mu }x}{\mu }\right)\right] + h.c =
\frac{4}{\sqrt{\pi}} \, \exp\left[\frac{(p+\tilde{\mu }x)^{2}}{\mu }\right]
\sum_{k=0}^{\infty} 
\mathcal{H}_{k-1}\left[i\,\mu ^{-{1}/{2}} (p+\tilde{\mu }x)\right]\,\frac{(\mu \, x^{2})^{k+\frac{1}{2}}}{(2k+1)!},
\end{eqnarray}
which leads to
\begin{eqnarray}
\label{finalform2}
W^{\alpha}(x,\,p;\,\tau_{})
&=& \frac{4}{\pi}\frac{\Gamma(3/2+\alpha)}{\Gamma(1+\alpha)}
(\mu \,x^2)^{1+\alpha}
\exp\left(-\mu \,x^2\right)\nonumber\\
&& \,
\sum_{k=0}^{\frac{1}{2}+\alpha}\frac{x^{-2k}}{\Gamma(1+k)\,\Gamma(3/2+k+\alpha)}\,
\frac{d^k}{d\mu ^k}
\sum_{q=0}^{\infty} 
\mathcal{H}_{q-1}\left[i\,\mu ^{-{1}/{2}} (p+\tilde{\mu }x)\right]\,\frac{(\mu \, x^{2})^{q}}{(2q+1)!}. \qquad
\label{eqn29}
\end{eqnarray}}.

\subsection{Coincident classical trajectories}

Considering the phase-space element $d\Pi_a\,da \equiv \,dp\,dx$ and the associated Poisson brackets given by $\{a,\,\Pi_a\}_{\mbox{\tiny PB}} \equiv \{x,\, p\}_{\mbox{\tiny PB}} = 1$, the dynamics of the Hamiltonian associated to Eq.~(\ref{eqn16}), for $\ell=0$, is given by the corresponding Hamilton equations for $x$ and $p$ assuming that the time dependence is driven by $H = H_{\mathcal{C}} - E$, with
\begin{equation}
H(x,p) = \frac{g_{_{C}}^{{1}/{2}}}{2} \left[p^2+ x^{2}+ \frac{4 \alpha^2 -1}{4 x^{2}} - 2 \alpha\right].
\label{eqn40}
\end{equation}
The classical trajectories are characterized by the condition $H_{\mathcal{C}}=E$, and
\begin{equation}
p_{_{\mathcal{C}}}=\pm \left[2\Delta -\left(x_{_{\mathcal{C}}}^{2}+ \frac{4 \alpha^2 -1}{4 x_{_{\mathcal{C}}}^{2}}\right)\right]^{1/2},\qquad\mbox{with}\quad \Delta = \alpha+{g_{_{C}}^{-{1}/{2}}}{E},
\label{eqn41}
\end{equation}
where $E$ is the classical energy and the index ``${{\mathcal{C}}}$'' has been introduced to denote classical quantities.
The evaluation of the Poisson brackets yields
\begin{eqnarray}
\dot{p}_{_{\mathcal{C}}} &=& \,\{p_{_{\mathcal{C}}},\,H\}_{\mbox{\tiny PB}} = - g_{_{C}}^{{1}/{2}}\,\left(x_{_{\mathcal{C}}}^{2}+ \frac{4 \alpha^2 -1}{4 x_{_{\mathcal{C}}}^{2}}\right),\\
 \dot{x}_{_{\mathcal{C}}} &=& \,\{x_{_{\mathcal{C}}},\,H\}_{\mbox{\tiny PB}} = + g_{_{C}}^{{1}/{2}}\,p_{_{\mathcal{C}}},
\label{eqn42}
\end{eqnarray}
where ``{\em dots}'' denote derivatives with respect to the classical time.
Through the constraint Eqs.~(\ref{eqn41}) and (\ref{eqn42}) one has
\begin{eqnarray}
\dot{x}_{_{\mathcal{C}}} &=&\pm g_{_{C}}^{{1}/{2}}\,\left[2\Delta -\left(x^{2}+ \frac{4 \alpha^2 -1}{4 x^{2}}\right)\right]^{{1}/{2}},
\label{eqn43}
\end{eqnarray}
which can be rewritten in terms of $\eta = x_{_{\mathcal{C}}}^2$ as
\begin{eqnarray}
\dot{\eta} &=&\pm g_{_{C}}^{{1}/{2}}\,\left[\frac{1- 4 \alpha^2}{4} + 2\eta\Delta - \eta^2\right]^{{1}/{2}}.
\label{eqn44}
\end{eqnarray}
By identifying the time variable with $\tau$ from the quantum framework, classical and quantum dynamics can be compared. Solving Eq.~(\ref{eqn44}), one obtains
\begin{eqnarray}
\eta_{\mp}(\tau) \equiv \eta_{\mp}\bb{\tau_{}} &=& \Delta \mp \varkappa^{{1}/{2}}\sin(\vartheta+ \omega \tau_{}),
\label{eqn45}
\end{eqnarray}
with $\vartheta$ arbitrary and $$ \varkappa = \frac{1}{4} +{2g_{_{C}}^{-{1}/{2}}\alpha E}+{g_{_{C}}^{-1}E^2}.$$ 

\begin{figure}
\includegraphics[scale=0.56]{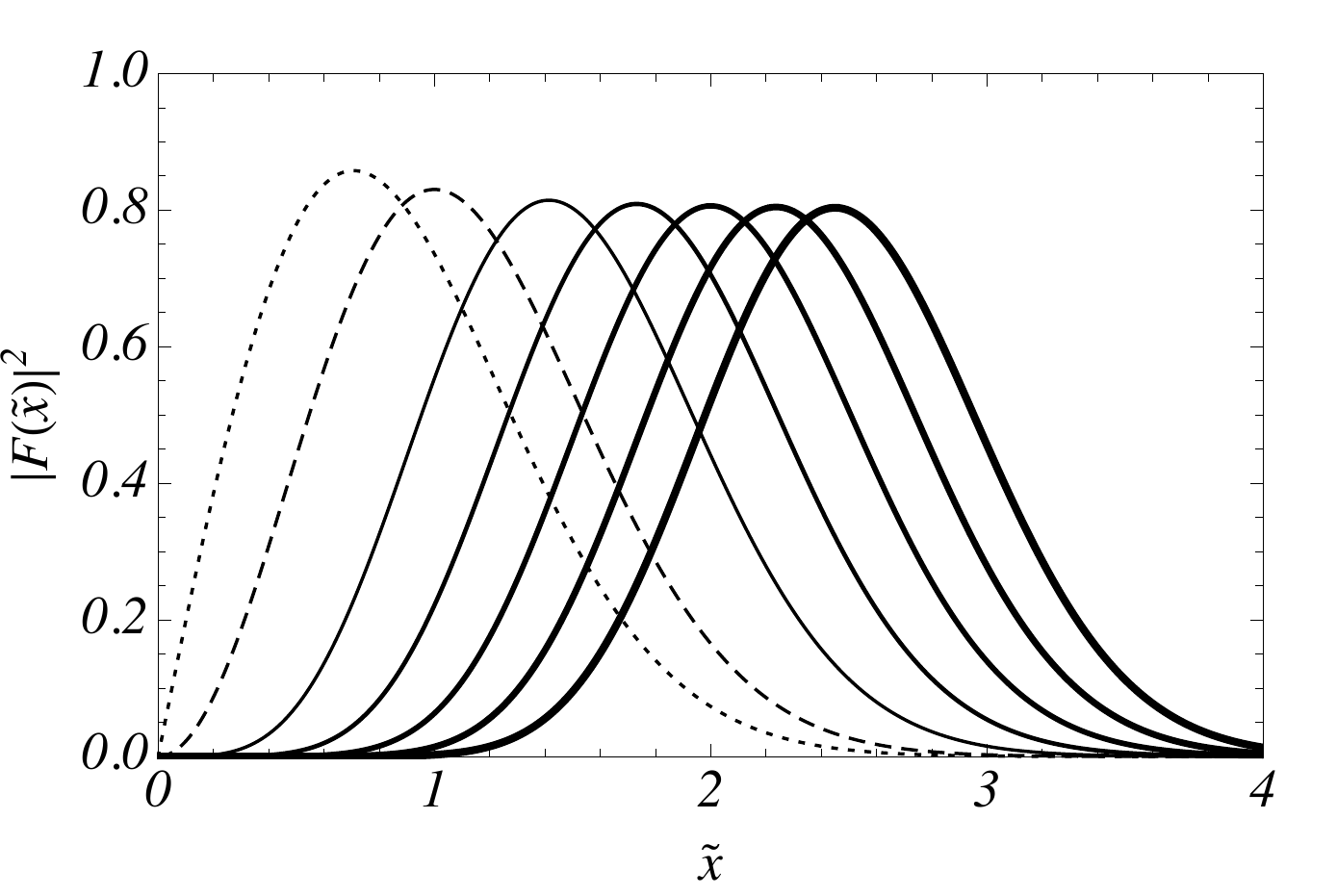}
\includegraphics[scale=0.56]{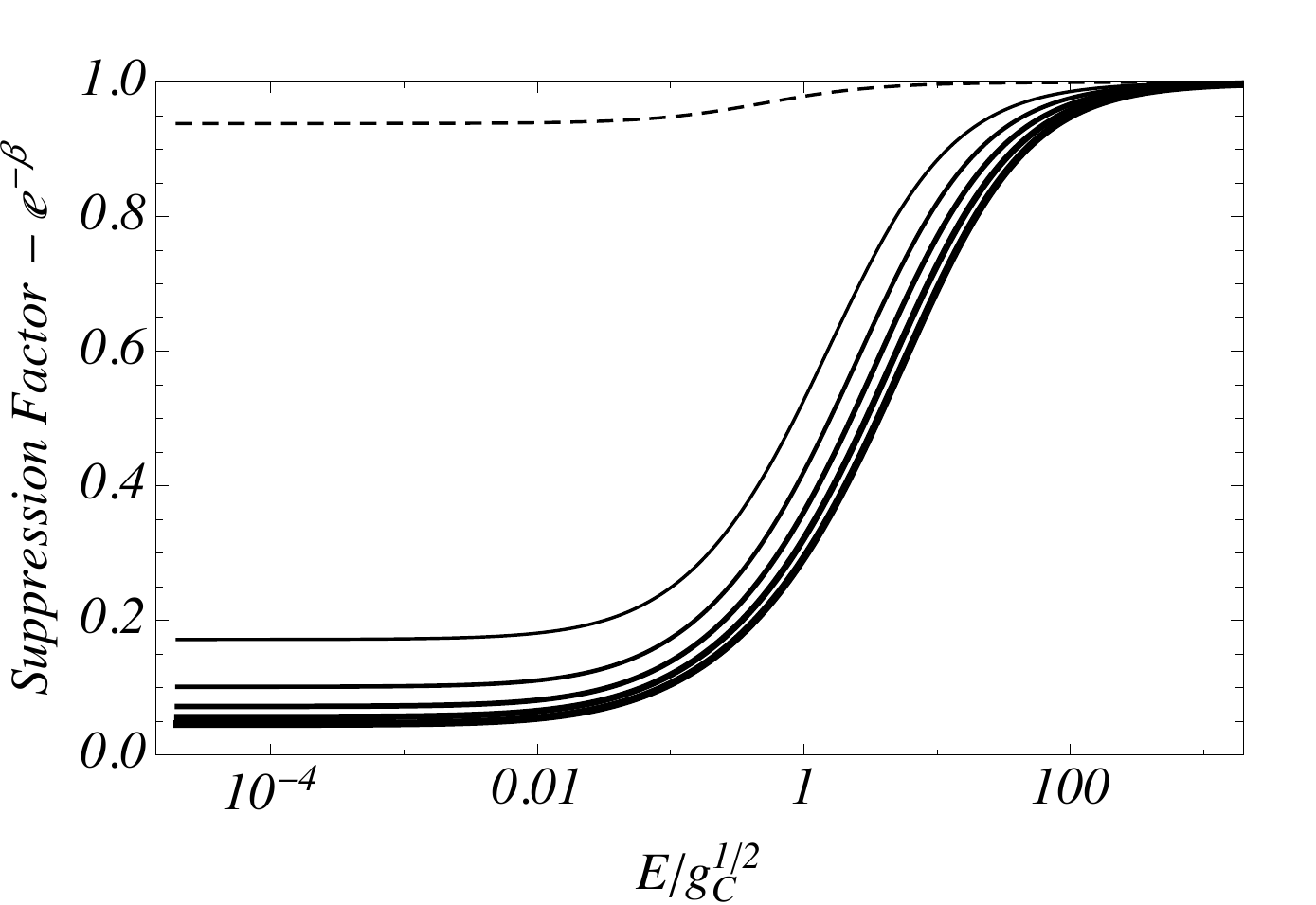}
\renewcommand{\baselinestretch}{.85}
\caption{
(First plot) Comoving modulus of the quasi-Gaussian wave function, $F_{\alpha}(\tilde{x})$, for the comoving coordinate $\tilde{x}$. (Second plot) Suppression factor, $\exp(-\beta)$, for the energy parameter $E/g_{_{C}}^{{1}/{2}}$.
The plots are for $\alpha = 0$ (dotted lines), $1/2$ (dashed lines), $3/2$ (thinest solid lines), $5/2$, $7/2$, $9/2$ and $11/2$ (thickest solid lines).}
\label{Figura002}
\end{figure}

For $\vartheta= \pi/2$ one has the classical solution
\begin{eqnarray}
x_{_{\mathcal{C}}\mp}^2\bb{\tau_{}} &=& \Delta \mp \varkappa^{{1}/{2}}\cos(\omega \tau_{}),
\label{eqn46}
\end{eqnarray}
which constrains the arguments of the Wigner function, $W^{\alpha}(x,\,p;\, \tau_{})$, to time independent values, i.e. the {\em stationary profile} of $W^{\alpha}(x_{_{\mathcal{C}}}\bb{\tau_{}},\,p_{_{\mathcal{C}}}\bb{\tau_{}};\, \tau_{})$ along the classical trajectory guarantees that $W^{\alpha}(x,\,p;\, \tau_{})$ returns time-dependent averaged values of the quantum observables, $x$ and $p$, that match the classical results, likewise to what happens with a Gaussian function for the harmonic oscillator problem.
This can be verified by observing the dynamical behavior of the arguments of $W^{\alpha}(x_{_{\mathcal{C}}}\bb{\tau_{}},\,p_{_{\mathcal{C}}}\bb{\tau_{}};\, \tau_{})$,
\begin{equation}
\tilde{x}^2_{(\beta,\tau_{})} = \mu_{(\beta,\tau_{})} x_{_{\mathcal{C}}}^2\bb{\tau_{}} = 
\frac{\sinh(\beta)\left(\Delta - \varkappa^{{1}/{2}}\cos(\omega\tau_{})\right)}{\cosh(\beta) - \cos(\omega\tau_{})} \equiv \left(\alpha^2-\frac{1}{4}\right)^{{1}/{2}},
\label{eqn24B}
\end{equation}
where $\mu$ is given by Eq.~\eqref{eqn23} and one constrains the $\beta$ parameter to 
\begin{equation}
\beta = \mbox{arctanh}\left[\frac{(4\alpha^2-1)^{{1}/{2}}}{2(\alpha +g_{_{C}}^{-{1}/{2}}E)}\right],
\label{eqn50}
\end{equation}
which is depicted in the second plot of Fig.~\ref{Figura002}.
From Eqs.(\ref{eqn23}) and (\ref{tilde}) one also notices that $\dot{\mu } = \tilde{\mu }\mu \omega$ and therefore\footnote{In fact, after simple manipulations one shows that $x\,p +\tilde{\mu }\,x^2 = 0$ for the above choice of parameters.}
\begin{equation}
\frac{d}{d\tau_{}}(x\,p +\tilde{\mu }\,x^2) = (\mu \omega)^{-1} \frac{d}{d\tau_{}}(\mu \,x^2) = 0.
\label{eqn50B}
\end{equation}
Eq.(\ref{eqn50B}) gives the exact constraint on the quantum mechanical superposition that corresponds to the quantum state $W^{\alpha}(x_{_{\mathcal{C}}}\bb{\tau_{}},\,p_{_{\mathcal{C}}}\bb{\tau_{}};\, \tau_{})$, which, in its turn, reproduces the classical profile of the evolution of the scale parameter, $a$: the classical cosmological solutions can be traced back to its quantum origin.

\section{HL Wigner flow and phase-space local quantum distortions}

For the phase-space described in terms of the coordinates $(a,\Pi_a)$, the conservation of probabilities is translated into the continuity equation (cf. Eq.~(\ref{eqn51}) from Appendix II),
\begin{eqnarray}
\frac{\partial W^{\alpha}}{\partial {t}} + \frac{\partial J_a^{\alpha}}{\partial a}+\frac{\partial J_{\Pi_a}^{\alpha}}{\partial \Pi_a} &=&
\frac{\partial W^{\alpha}}{\partial t} + 2 g_{_{C}}^{{1}/{2}} \left[\frac{\partial}{\partial x}\left(\frac{J_a^{\alpha}}{2g_{_{C}}^{{1}/{4}}}\right) + \frac{\partial}{\partial p}\left(\frac{J_{\Pi_a}^{\alpha}}{2g_{_{C}}^{{3}/{4}}}\right)\right]\nonumber\\
&=&
\frac{1}{\omega}\frac{\partial W^{\alpha}}{\partial \tau_{}} + \frac{\partial J_x^{\alpha}}{\partial x} + \frac{\partial J_p^{\alpha}}{\partial p}\nonumber\\&=&0,
\label{eqn66}
\end{eqnarray}
where again it has been set that $t\equiv \tau$, with $2 g_{_{C}}^{{1}/{2}}= \omega$, in order to recast the Wigner current components as
\begin{eqnarray}
J^{\alpha}_x(x,\,p;\,\tau_{})&=& (2g_{_{C}}^{{1}/{4}})^{-1} J_a^{\alpha} = (2g_{_{C}}^{{1}/{4}})^{-1} \,\Pi_a \,W^{\alpha}(a,\,\Pi_a;\, \tau_{}) = \frac{p}{2}\,W^{\alpha}(x,\,p;\, \tau_{}),
\label{eqn501}
\\
J^{\alpha}_p(x,\,p;\,\tau_{})
&=& (2g_{_{C}}^{{3}/{4}})^{-1} J_{\Pi_a}^{\alpha}\nonumber\\
&=& - (2g_{_{C}}^{{3}/{4}})^{-1}
\sum_{k=0}^{\infty} \left(\frac{i}{2}\right)^{2k}\frac{1}{(2k+1)!} \left[\left(\frac{\partial~}{\partial a}\right)^{2k+1}\hspace{-.5 cm}V(a)\right]\,\left(\frac{\partial~}{\partial \Pi_a}\right)^{2k}\hspace{-.3 cm}W^{\alpha}(a,\,\Pi_a;\, \tau_{})\nonumber\\
&=& -\frac{1}{2}\sum_{k=0}^{\infty} \left(\frac{i}{2}\right)^{2k}\frac{1}{(2k+1)!} \, \left[\left(\frac{\partial~}{\partial x}\right)^{2k+1}\hspace{-.5 cm}\mathcal{V}(x)\right]\,\left(\frac{\partial~}{\partial p}\right)^{2k}\hspace{-.3 cm}W^{\alpha}(x,\,p;\, \tau_{}).
\label{eqn502}
\end{eqnarray}
where $\mathcal{V}(x)$ is given by Eq.~(\ref{eqn17}), and the overall multiplying factor (an one-half factor in the final form for this case) is not relevant and can be absorbed by the normalization definitions of the parameter $\omega$.
The above phase-space vectors follow the fluid equations from Appendix II (for a canonical coordinates obeying $[\xi_x,\,\xi_p]=i\hbar$ converted into $[a,\,\Pi_a] = i$) through a generalization of the Wigner flow formalism to the WdW quantum cosmological approach.

Notice that by truncating the above sum at $k=0$ one recovers the classical results.
Quantum and non-linear corrections arise from the contributions due to the infinite expansion,
\begin{eqnarray}
 -\frac{1}{2}\sum_{k=1}^{\infty} \left(\frac{i}{2}\right)^{2k}\frac{1}{(2k+1)!} \, \left[\left(\frac{\partial~}{\partial x}\right)^{2k+1}\hspace{-.5 cm}\mathcal{V}(x)\right]\,\left(\frac{\partial~}{\partial p}\right)^{2k} \hspace{-.3 cm}W^{\alpha}(x,\,p;\, \tau_{}),
\label{eqn503}
\end{eqnarray}
which, for $\mathcal{V}(x)$, can be accurately described through an analytical expression.
The constant and the harmonic oscillator contribution from $\mathcal{V}(x)$ are easy to manipulate and do not introduce any kind of quantum back-flow \cite{QBR}. Since one has the potential proportional to $x^2$, the harmonic oscillator Wigner current contribution is simply given by 
\begin{eqnarray}
J^{\alpha(HO)}_p(x,\,p;\,\tau_{})
&=& -\frac{x}{2}\,W^{\alpha}(x,\,p;\, \tau_{}),
\label{eqn502B}
\end{eqnarray}
i.e. the amplitude for classical and quantum cases are the same.

Otherwise, one should pay some attention to the contribution due to the potential term which is proportional to $1/x^{2}$ (cf. Eq.~(\ref{eqn17})) and, eventually, to perturbative contributions due to $\ell\,x^4$ (cf. Sec. V)).

As to perform the analytical calculations, one first notices that 
\begin{equation}
\left(\frac{\partial~}{\partial x}\right)^{2k+1}\frac{1}{x^2} = -(2k+2)\frac{(2k+1)!}{x^{2k+3}}
\label{eqn68},
\end{equation}
and
\begin{equation}
\left(\frac{\partial~}{\partial p}\right)^{2k}W^{\alpha}(x,\,p;\, \tau_{}) =
\frac{1}{\pi}\int_{-\infty}^{+\infty}dy\,(2\,i\,y)^{2k}\,\exp(2\,i\,p\,y)\,\mathrm{F}^*_{\alpha}(x+y)\,\mathrm{F}_{\alpha}(x-y).
\label{eqn69}
\end{equation}
One can then work out the sum in Eq.~(\ref{eqn503}) related to the term proportional to $1/x^{2}$ in $\mathcal{V}(x)$ as to have
\begin{eqnarray}
\lefteqn{-\frac{1}{2}\sum_{k=1}^{\infty} 
\left(\frac{i}{2}\right)^{2k}\frac{1}{(2k+1)!} 
\left[\left(\frac{\partial~}{\partial x}\right)^{2k+1}
\frac{1}{x^2}\right]\,\left(\frac{\partial~}{\partial p}\right)^{2k}\, W^{\alpha}(x,\,p;\, \tau_{})=}\nonumber\\
&&= \frac{1}{x^3} 
\int_{-\infty}^{+\infty}dy\,\sum_{k=1}^{\infty} (-1)^{2k}
\frac{(2k+1)!}{(2k+1)!}(k+1)
\left(\frac{y}{x}\right)^{2k} \,\exp(2\,i\,p\,y)\,\mathrm{F}^*_{\alpha}(x+y)\,\mathrm{F}_{\alpha}(x-y)
\nonumber\\
&&= \frac{1}{\pi x^3}\int_{-\infty}^{+\infty}dy\,\frac{d~}{d\varepsilon}
\left(\sum_{k=1}^{\infty} \varepsilon^{k+1}\right)
\,\exp(2\,i\,p\,y)\,\mathrm{F}^*_{\alpha}(x+y)\,\mathrm{F}_{\alpha}(x-y)\nonumber\\
&&= \frac{x}{\pi}\int_{-\infty}^{+\infty}dy\,(x^2-y^2)^{-2}\,\exp(2\,i\,p\,y) 
\,\mathrm{F}^*_{\alpha}(x+y)\,\mathrm{F}_{\alpha}(x-y)\nonumber\\
&&= x \mu ^{2}\,\frac{\Gamma(\alpha-1)}{\Gamma(\alpha+1)}
W^{\alpha-2}(x,\,p;\, \tau_{})
\label{eqn504},
\end{eqnarray}
where $\varepsilon = y^2/x^2 < 1$ has been considered, and without loss of generality, due to the Heaviside distribution in each function $\mathrm{F}$, since $y\in [-x,x]$ then
$$\frac{d~}{d\varepsilon}\left(\sum_{k=1}^{\infty} \varepsilon^{k+1}\right) = (1-\varepsilon)^{-2}.$$

The contribution from Eq.~(\ref{eqn504}) must be multiplied by $(1-4\alpha^2)/8$ as to match the contribution from $\mathcal{V}(x)$ which, once it is added to the contribution from Eq.~(\ref{eqn502B}), results into
\begin{eqnarray}
J^{\alpha}_p(x,\,p;\,\tau_{})
&=& -\frac{x}{2}\left(
W^{\alpha}(x,\,p;\, \tau_{}) + \frac{1-4\alpha^2}{4}\, \mu ^{2}\,\frac{\Gamma(\alpha-1)}{\Gamma(\alpha+1)} W^{\alpha-2}(x,\,p;\, \tau_{})\right),
\label{eqn505}
\end{eqnarray}
from which one obtains
\begin{eqnarray}
J^{\alpha(Cl)}_p(x,\,p;\,\tau_{})
&=& - \frac{1}{2}W^{\alpha}(x,\,p;\, \tau_{})\left(x + \frac{1-4\alpha^2}{4x^3}\right),
\label{eqn505B}
\end{eqnarray}
for the classical limit. 

\begin{figure}
\includegraphics[scale=0.45]{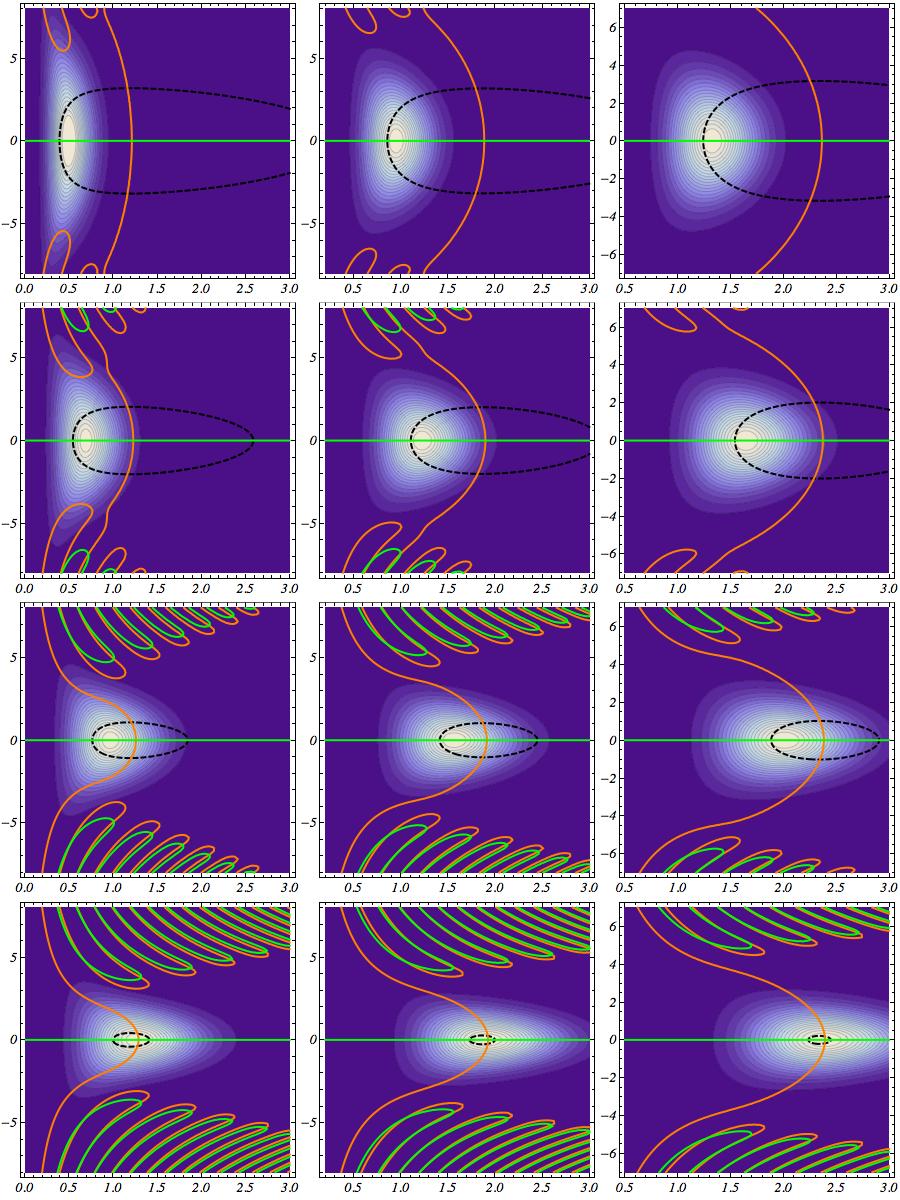}
\renewcommand{\baselinestretch}{.85}
\caption{
(Color online) Features of the Wigner flow for the quasi-Gaussian wave function, $W^{\alpha}(x,\,p;\,\tau_{})$, in the $x - p$ plane, at $\tau_{}=0$ .
Green contour lines are for $J^{\alpha}_x(x,\,p;\, 0) = 0$ and orange contour lines are for $J^{\alpha}_p(x,\,p;\, 0) = 0$.
Green (orange) contour lines are bounds for the reversal of the Wigner flow in the $x(p)$ direction.
The plots are for $E = 5,\, 2,\, 0.5$, and $0$, from top to bottom, and for $\alpha = 3/2,\,7/2$, and $11/2$, from left to right.
The color scheme background shows the Wigner function profiles of $W^{\alpha}$ for $\omega\tau_{}=0$, with the details of the domains quantum fluctuations bounded by green and orange lines.
}
\label{Figura004}
\end{figure}

\begin{figure}
\includegraphics[scale=0.58]{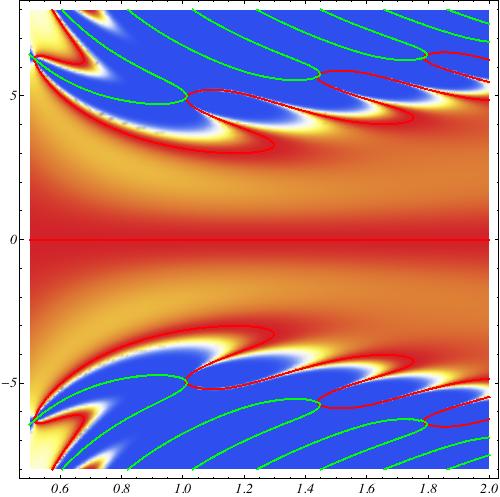}
\includegraphics[scale=0.58]{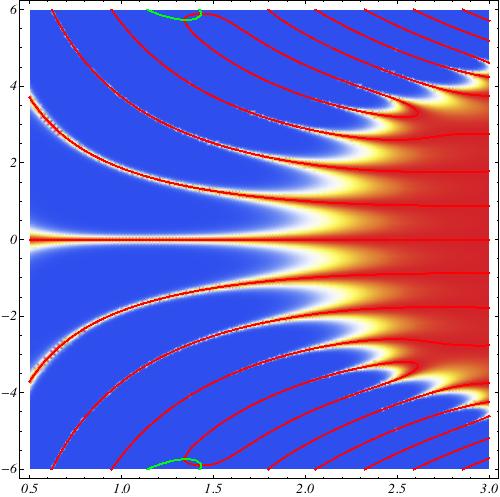}
\renewcommand{\baselinestretch}{.85}
\caption{
(Color online) Behavior of the Liouvillian quantifier parameterized by $\mbox{sech}(\mbox{\boldmath $\nabla$}_{\xi} \cdot \mathbf{u})$ for the quasi-Gaussian Wigner function, $W^{\alpha}(x,\,p;\,\tau_{})$ in the phase-space ($x - p$ plane). Red-lines are for $\mbox{\boldmath $\nabla$}_{\xi} \cdot \mathbf{u} = 0$ and the {\em TemperatureMap} color scheme (from blue-regions, $\mbox{sech}(\mbox{\boldmath $\nabla$}_{\xi} \cdot \mathbf{u}) \sim 0$, to red-regions, $\mbox{sech}(\mbox{\boldmath $\nabla$}_{\xi} \cdot \mathbf{u}) \sim 1$) reinforces the approximated Liouvillian behavior for red-regions. Green-lines mark the zeros of the coincident values of $W^{\alpha}(x,\,p;\, 0) = J^{\alpha}_x(x,\,p;\, 0)$, where the $\mbox{\boldmath $\nabla$}_{\xi} \cdot \mathbf{u}$ becomes unbounded (maximal non-Liouvillian behavior). The plots are for $\alpha= 3/2$ (top) and $11/2$ (bottom).
}
\label{Figura005}
\end{figure}

In fact, for $\tau_{}=0$, quantum fluctuations are highly suppressed by the quasi-Gaussian profile of the Wigner function, as it has been identified by the light-dark color scheme in Fig.~(\ref{Figura004}).
The Wigner flow stagnation points are defined by orange-green crossing lines, where $J_x^{\alpha} = J_p^{\alpha}=0$, so that quantum features are completely suppressed for the series expansion, Eq.~(\ref{eqn503}), truncated at $k=0$.
The classical trajectory portrait is shown as a collection of black dashed lines.
The non-Liouvillian behavior \cite{Steuernagel3} is depicted in Fig.~(\ref{Figura005}) for arbitrary choices of $\alpha$.

Thus, by comparing expressions for the classical and quantum Wigner currents, one is able to quantify the quantum fluctuations due to $\Delta J^{\alpha}_p(x,\,p;\,\tau_{}) = J^{\alpha}_p(x,\,p;\,\tau_{}) - J^{\alpha(Cl)}_p(x,\,p;\,\tau_{})$ over any specific volume $\Delta p\, \Delta x$ of the phase-space.

\subsubsection*{Classicality}

Given that $W^{\alpha}(x,\,p;\, \tau_{})$ corresponds to a pure state, the global fluid behavior sets vanishing values for the rate of change of purity \cite{01A,02A}, $\dot{\mathcal{P}}$ (see the Appendix I).
Such a behavior is interpreted as due to phase space symmetry and closure properties \cite{01A,02A}.
Therefore, only local quantum fluctuations can be identified and quantified as a measure of non-classicality for the Wigner flow \cite{MeuPaper}.

For a periodic motion defined by closed trajectories as, for instance, those obtained from Eqs.~(\ref{eqn42})-(\ref{eqn44}), one can associate the classical trajectory to the two-dimensional boundary surface from the Wigner flow. 

According to the results discussed in Appendix II \cite{MeuPaper}, the local features of non-classicality for periodic motions defined by classical trajectories (cf. Eqs.~(\ref{eqn42})-(\ref{eqn44})) can be quantified in terms of an integrated periodic probability flux enclosed by the classical surface, $\mathcal{C}$, given by (cf. Eqs.~(\ref{eqn51BB})-(\ref{eqn51EE}))
\begin{equation}
\frac{D~}{D\tau_{}}\mbox{Prob}_{(\mathcal{C})}
\bigg{\vert}_{{\tau_{}} = \frac{2\pi}{\omega} }
= -
\int_{0}^{\frac{2\pi}{\omega}} d\tau_{}\, \Delta J^{\alpha}_p(x_{_{\mathcal{C}}}\bb{\tau_{}},\,p_{_{\mathcal{C}}}\bb{\tau_{}};\tau_{})\,\,p_{_{\mathcal{C}}}\bb{\tau_{}}.
\label{eqn51MM}
\end{equation}

From Eqs.~\eqref{eqn505} and \eqref{eqn505B}, one writes 
\small\begin{eqnarray}
\Delta J^{\alpha}_p(x_{_{\mathcal{C}}}\bb{\tau_{}},\,p_{_{\mathcal{C}}}\bb{\tau_{}};\tau_{}) &=& 
\frac{1-4\alpha^2}{8} \bigg{(} x_{_{\mathcal{C}}}\bb{\tau_{}}\,\mu^{2} \,\frac{\Gamma(\alpha-1)}{\Gamma(\alpha+1)} W^{\alpha-2}(x_{_{\mathcal{C}}}\bb{\tau_{}},\,p_{_{\mathcal{C}}}\bb{\tau_{}};\, \tau_{})\qquad\qquad\nonumber\\
&&\quad\qquad\qquad\qquad\qquad\qquad +
x^{-3}_{_{\mathcal{C}}}\bb{\tau_{}}\, W^{\alpha}(x_{_{\mathcal{C}}}\bb{\tau_{}},\,p_{_{\mathcal{C}}}\bb{\tau_{}};\, \tau_{}) \bigg{)},
\label{eqn505BBB1}
\end{eqnarray}\normalsize
with $x_{_{\mathcal{C}}}^2\bb{\tau_{}}$ given by Eq.~\eqref{eqn46},
and $p_{_{\mathcal{C}}}^2\bb{\tau_{}}$ obtained from Eq.~\eqref{eqn41}.
In fact, according to the results from Eqs.~\eqref{eqn41}-\eqref{eqn46} the above integral can be evaluated by noticing the stationary behavior of the multiplying factor
\begin{eqnarray}
S^{\alpha}\equiv S^{\alpha}(x_{_{\mathcal{C}}}\bb{\tau_{}},\,p_{_{\mathcal{C}}}\bb{\tau_{}};\tau_{}) &=& 
\frac{1-4\alpha^2}{8} \bigg{(} x^{4}\bb{\tau_{}}\,\mu ^{2} \,\frac{\Gamma(\alpha-1)}{\Gamma(\alpha+1)} W^{\alpha-2}(x_{_{\mathcal{C}}}\bb{\tau_{}},\,p_{_{\mathcal{C}}}\bb{\tau_{}};\, \tau_{})\nonumber\\
&&\qquad\qquad\qquad\qquad + W^{\alpha}(x_{_{\mathcal{C}}}\bb{\tau_{}},\,p_{_{\mathcal{C}}}\bb{\tau_{}};\, \tau_{}) \bigg{)},
\label{eqn505BBB2}
\end{eqnarray}
with $dS^{\alpha}/d \tau_{} = 0$, which sets
\begin{eqnarray}
\frac{D~}{D\tau_{}}\mbox{Prob}_{(\mathcal{C})}
\bigg{\vert}_{\tau_{} = \frac{2\pi}{\omega}} 
&=& - S^{\alpha}
\int_{0}^{\frac{2\pi}{\omega}} d\tau_{}\,
\,p_{_{\mathcal{C}}}\bb{\tau_{}}\,x^{-3}_{_{\mathcal{C}}}\bb{\tau_{}}\nonumber\\
&=& - S^{\alpha}
\int_{0}^{\frac{2\pi}{\omega}} d\tau_{}\,
\,\varkappa^{{1}/{2}}\,\sin(\omega \tau_{})\left(\Delta - \varkappa^{{1}/{2}}\,\cos(\omega \tau_{}) \right) ^{-4}\nonumber\\
&=& \frac{3}{\omega} S^{\alpha}
\int_{0}^{\frac{2\pi}{\omega}} d\tau_{}\,
\,\left(\Delta - \varkappa^{{1}/{2}}\,\cos(\omega \tau_{}) \right) ^{-3}\bigg{\vert}^{\tau_{} = \frac{2\pi}{\omega}}_{\tau_{}=0} \nonumber\\
&=& 0.
\label{eqn51SS2}
\end{eqnarray}

Given that, due to the quantum fluctuations, the integrand is non-vanishing along the parametric classical trajectory, the above result is quite auspicious as it depicts the quantum to classical transition of the HL scenario here described.

A second approach to quantify the quantum fluctuations can be established by the averaged value of $\Delta J^{\alpha}_p(x,\,p;\,\tau_{})$ over all the phase-space volume,
\begin{eqnarray}
\lefteqn{\Delta^{\alpha}\bb{\tau_{}} =\int_{0}^{\infty}\hspace{-.3 cm}dx\int_{-\infty}^{+\infty}\hspace{-.3 cm}dp \,W^{\alpha}\,\Delta J^{\alpha}_p(x,\,p;\,\tau_{})=}\nonumber\\&& 
\frac{4\alpha^{2}-1}{2}\frac{\mu ^{2(1+\alpha)}}{\pi^2\,\Gamma^2(1+\alpha)}\,\int_{0}^{\infty}\hspace{-.3 cm}dx \,x^{1+4\alpha}\,\exp\left(-2\,\mu x^2\right)\,\int_{-\infty}^{+\infty}\hspace{-.3 cm}dp
\,\exp\left(2\,i\,x\,p \,(s+r) \right)\times\nonumber\\
&& \qquad\qquad\int_{-1}^{+1} ds\int_{-1}^{+1} dr\,[(1-r^2)(1-s^2)]^{\frac{1}{2}+\alpha}\left[(1-s^2)^{-2}-1\right]\times\, \nonumber\\
&& \qquad\qquad\qquad\qquad\exp\left(-\mu x^2\,(r^2+s^2)\right) \exp\left(2i\,\tilde{\mu }_{(\beta,\tau_{})} x^2\,(s+r)\right).
\label{eqn70}
\end{eqnarray}
By following the strategy suggested by Eqs.~(\ref{eqnA04})-(\ref{eqnA05}) from Appendix I, one obtains
\begin{eqnarray}
\Delta^{\alpha}\bb{\tau_{}} &=& 
\frac{4\alpha^{2}-1}{2}\frac{\mu ^{2(1+\alpha)}}{\pi\,\Gamma^2(1+\alpha)}\,\int_{0}^{\infty}\hspace{-.3 cm}dx \,x^{4\alpha}\,\exp\left(-2\,\mu x^2\,(1+s^2)\right)\times\nonumber\\
&& \qquad\qquad\qquad\qquad\qquad\qquad\int_{-1}^{+1} ds\,(1-s^2)^{1+2\alpha}\left[(1-s^2)^{-2}-1\right]\\
&=& 
\mu ^{\frac{3}{2}} 
\frac{4\alpha^{2}-1}{{2}^{\frac{5}{2}+2\alpha}\pi}
\frac{\Gamma(1/2 +2\alpha)}{\Gamma^2(1+\alpha)}
\int_{-1}^{+1} ds\,{(1+s^2)}^{\frac{1}{2}+2\alpha} \,(1-s^2)^{1+2\alpha}\left[(1-s^2)^{-2}-1\right],\nonumber
\label{eqn71}
\end{eqnarray}
where the periodic time dependence is factorized from the $\alpha$ dependent parameters.

\begin{figure}
\includegraphics[scale=0.88]{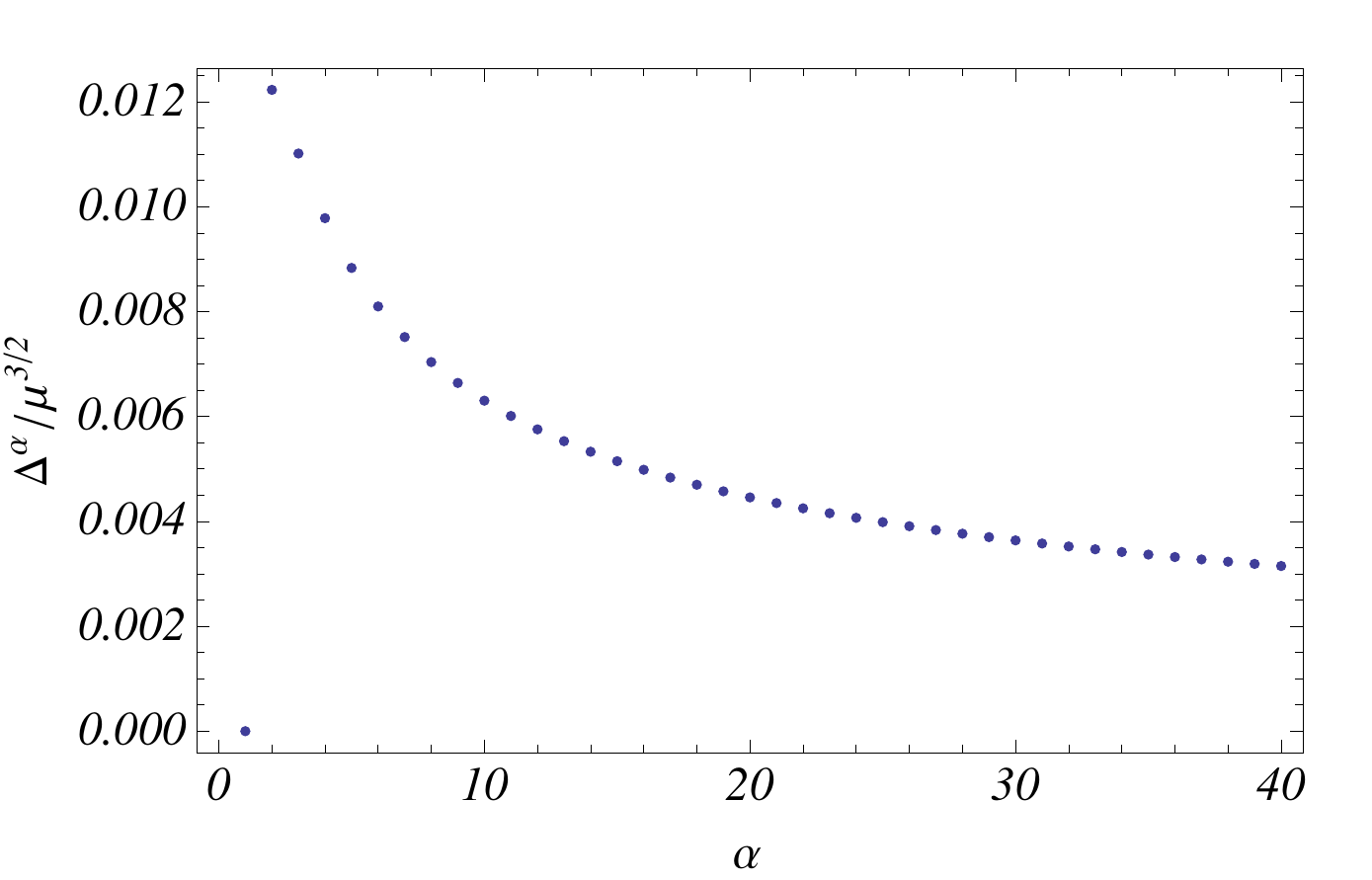}
\renewcommand{\baselinestretch}{.85}
\caption{
Quantum back-reaction quantifier, $\Delta^{\alpha}\bb{\tau_{}}$, normalized by the time dependence on $\mu ^{\frac{3}{2}}_{(\beta,\tau_{})}$, as a function of $\alpha$. Notice that for $\alpha =1/2$ one recovers the (classical) harmonic oscillator result.}
\label{Figura009}
\end{figure}

\begin{figure}
\vspace{-1 cm}\includegraphics[scale=0.31]{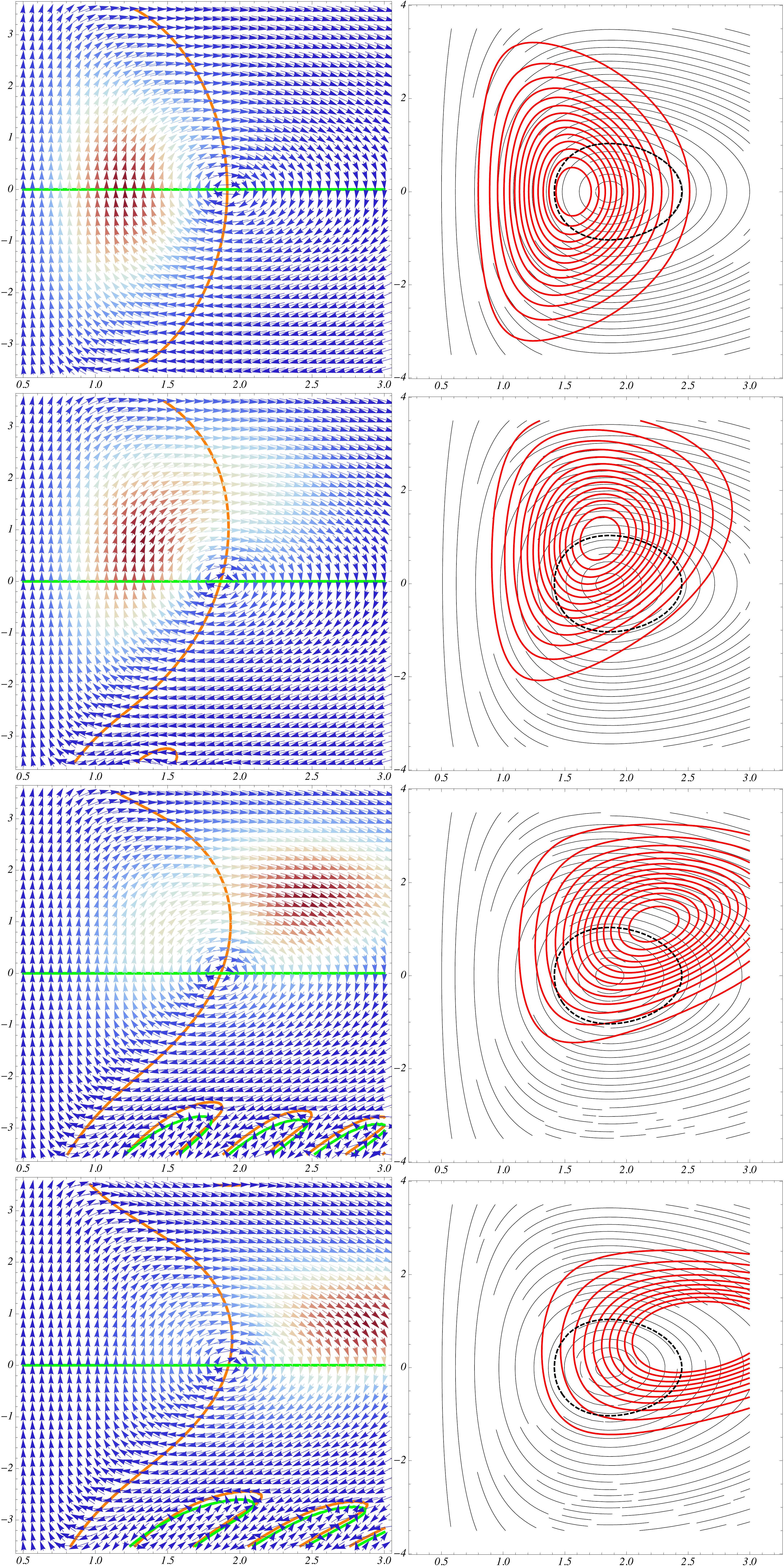}
\renewcommand{\baselinestretch}{.85}
\caption{
(Color online) 
First column:
Time evolution of the normalized quantum Wigner flow fields, $\mathbf{J}/\vert\mathbf{J}\vert$, scaled by the {\em ThermometerColor} scheme (from zero (blue) to one (red)) in the $x - p$ plane.
As before, the green (orange) contour lines are for $J^{\alpha}_{x(p)}(x,\,p;\, \tau_{}) = 0$.
Second column:
Time evolution of the dominant region (red contours) of the Wigner function, $W^{\alpha}(x,\,p;\, \tau_{})$ superimposed by classical trajectories (thin black streamlines).
The plots are for $\omega\tau_{} = 0$, $\pi/4$, $\pi/2$, and $3\pi/4$,
from top to bottom, with $E=1/2$ and $\alpha=7/2$.
}\label{Figura006}
\end{figure}

\begin{figure}
\vspace{-1 cm}\includegraphics[scale=0.4]{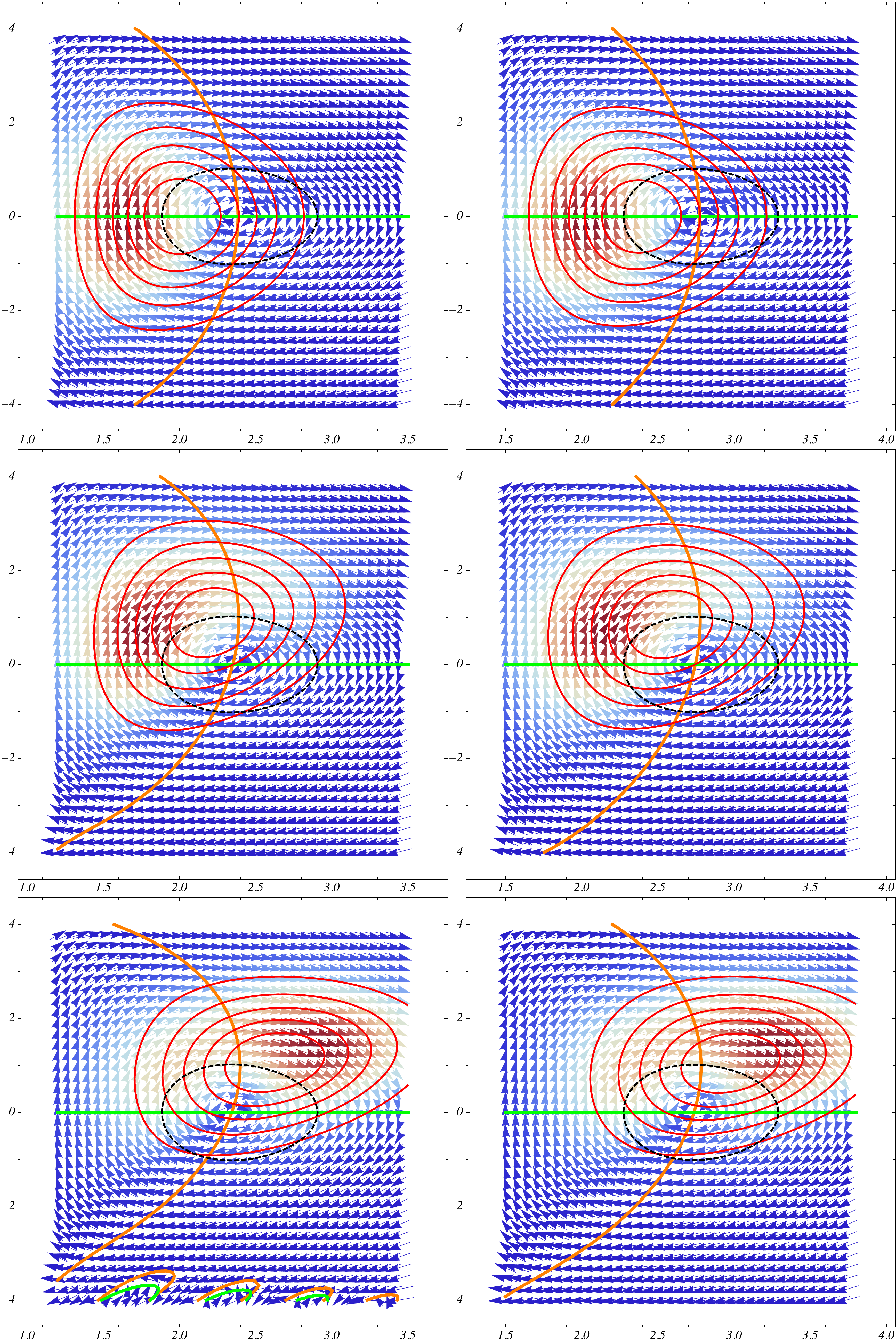}
\renewcommand{\baselinestretch}{.85}
\caption{
(Color online) Time evolution of the Wigner function, $W^{\alpha}(x,\,p;\, \tau_{})$, and the corresponding features of the Wigner flow in the phase-space ($x - p$ plane), for $\omega\tau_{} = 0$ (first row), $\pi/4$ (second row), and $\pi/2$ (third row) (similar to Fig.~\ref{Figura006}), and $E =1/2$ and $\alpha = 11/2$ (first column) and $\alpha = 15/2$ (second column).
The quantum flows are in agreement with Fig.~\ref{Figura006}: the average values (and not the maximal probability values) for the canonical coordinates follow the classical trajectory.
With respect to lower values of the parameter $\alpha$, the quantum effects are suppressed.}
\label{Figura007}
\end{figure}

The behavior of this quantity is shown in Fig.~\ref{Figura009}, where one notices that the influence of quantum fluctuations are suppressed by increasing values of $\alpha$, as it was qualitatively depicted in Fig.~\ref{Figura004}. The action from Eq.~(\ref{eqn10}) indeed yields an increasing value proportional to $\alpha$ as the $t$ integration is performed. This is consistent with the expectation that classicality corresponds to the action, Eq.~(\ref{eqn10}), $\mathcal{S} \gg 1$ (in Planck units).
The results from Fig.~\ref{Figura009} are complemented by the flux map of Figs.~\ref{Figura006} and \ref{Figura007}.
 
Turning back to the qualitative analysis provided by the integrated flux, one should notice that the evolution of the Wigner function fits accurately the classical trajectory, as predicted by the sequence of Eqs.~(\ref{eqnstart})-(\ref{eqn50B}) from Sec. III.

Once one has $\alpha - 2 \geq 1/2$ as to satisfy the condition for obtaining the analytical expression for $J^{\alpha}_p(x,\,p;\, \tau_{})$, the qualitative interpretation of the results from Figs.~\ref{Figura006} and \ref{Figura007} is not affected by the choice of $E$ and $\alpha$.
Quantitatively, the quantum distortion is suppressed by increasing values of $\alpha$, which corresponds to the transition of quantum trajectories into classical ones.

Notice that the quantum superposition described by $W^{\alpha}(x,\,p;\, \tau_{})$ follows the classical trajectory (thick black dashed-lines) in spite of the presence of quantum back-reaction effects (residual red arrows) due to $W^{\alpha-2}(x,\,p;\, \tau_{})$ along the $\hat{p}$ direction. 

The Wigner flow stagnation points, which are typical quantum features, are defined by orange-green crossing lines, with $J_x^{\alpha} = J_p^{\alpha}= 0$. In the quantum cosmology context, the stagnation points vanish for $M_{\mbox{\tiny Pl}}^{-1} \to 0$.
They are identified by clockwise and anti-clockwise vortices (winding number equals to $+1$ and $-1$), separatrix intersections and saddle flows (winding number equals to $0$).
As a compensating effect, the contra-flux fringes (delimited by green and orange lines) emerge to brake the retarded evolution of the quantum flux. 
The classical profile does not exhibit such an overall locally compensation phenomena.

Finally, given that in the computation of the Wigner currents the quantum effects are accompanied by the non-linear contributions of the quantum potential, the exact result, Eq.~(\ref{eqn505}), obtained from the infinite expansion from Eq.~(\ref{eqn503}), guarantees that quantum corrections have been accurately accounted for in the above analysis.

The consistence of the above results can be assessed by comparison with analyses of the decoherence and classical correlations on the retrieval of classical behavior from the wave function of the Universe \cite{Habib01}.
In principle, the investigation of quantum decoherence and of an emergence of classicality could be performed through a kind of wave packed spreading mechanism in the Schr\"odinger picture of quantum mechanics.
Considering that this is rigorously suitable for the evolution of pure states, a broader description of how the loss of information takes place into the quantum to classical transition is more properly achieved through a density matrix description, for instance, through the Wigner-Weyl formalism, which provides a suitable picture of the quantum to classical transition.
In this picture, decoherence takes place through a coarse-graining procedure \cite{Habib01,Ballentine}, which implies into the spreading of the Wigner functions (and associated coordinate probability densities), whereas coordinate and momentum classical correlations require a localized version of them \cite{Habib01}.
That is, quantum coherence is lost through a coarse-graining procedure \cite{Habib01,Habib02} where either the density matrix is averaged over the phase-space variables or some external environment effect is considered.
The decoherence triggers suppression of the correlation between coordinate and momentum, although they coexist as the quantum pattern is suppressed (but not destroyed) by sharply peaked quantum superpositions.
That is the case of the HL cosmological system described along this section.
The quasi-Gaussian sharpness of the resulting Wigner functions guarantees the correlation between averaged values of position and momenta, at the same time that the quantum pattern is still present in the adjacent quantum fringes, as it can be seen in Figs.~\ref{Figura004} and \ref{Figura005}. Hence, no decoherence effect takes place in this framework.

According to Hartle and Geroch prescriptions for quantum cosmology \cite{Hartle2,Geroch}, likewise in the quantum mechanics standard interpretation, peaks in the quantum (quasi) distribution function are equivalent to predictions in quantum cosmology and, for the HL coherent peaked superposition discussed here, the resulting quantum dynamics is consistent with the classical correlation between position and momentum in a phase-space description.

Of course, it is worth mentioning that our description is achievable thanks to the mathematical manipulability of the Weyl transformed of the associated Laguerre polynomials that describe the HL eigensystem. Pure states corresponding to sharply peaked quasi-Gaussian wave functions are built without additional constraints, and these can be compared with those resulting from coarse-graining methods, required to set up the decoherence process. Given that no decoherence mechanism takes place in our framework, the quantum to classical transition is highly constrained by the choice of the HL cosmology, namely by the analytical manipulability provided by the WdW wave functions, and by the corresponding Wigner functions. 
In comparison, the WKB approximation and the coarse-graining procedure suggested in Ref.~\cite{Habib01} is more general. In fact, a crucial point for the accuracy of the WKB analysis \cite{Habib01} is the significance of the $\mathcal{O}(\hbar)$ quantum corrections. This is irrelevant at our scenario of HL cosmology where all orders in $\hbar$ are absorbed in our procedure (cf. Eqs.~(\ref{finalform}), (\ref{finalform2}), and (\ref{eqn504})).
In the WKB approximation, quantum corrections are suppressed even though quantum interference fringes yield a Wigner function with a very large number of peaks which average to zero. In quantum cosmology, these cannot be neglected as their contribution yields observable averaged values. Therefore, in some cases, it suppresses the classical correlations. This is not the case at our HL approach, where the quasi-Gaussian quantum superposition in the phase-space guarantees the effectiveness of a WKB analysis\footnote{In particular, for the HL associated classical dynamics driven by the Hamiltonian Eq.~(\ref{eqn40}), the classical time variable is identified with $\tau$ from the quantum framework, as to guarantee the reproduction of classical trajectories by the HL wave function quantum superpositions from Eqs.~(\ref{supp})-(\ref{eqnstart}). Such a correspondence provides the elements for identifying $\tau$ with the WKB semiclassical time \cite{VilenBerto1} in a generalized analysis where (cf. Ref.~\cite{Ballentine}, pag. 402) the classical probability density is proportional to the time that the particle spends in an interval $\Delta x$ such that the coarse-grained quantal probability density agrees with the classical probability density.}. However, the same cannot be asserted, for instance, to the following discussion of an analogous bounce model where, as one shall see, the presence of quantum interference effects is only completely captured by exact Wigner functions.

\subsubsection*{Wigner flow and quantum effects for bounce models}

One of the fundamental questions in quantum cosmology concerns the initial singularity. 
Once the quantum cosmology in the minisuperspace framework admits a universe described by a wave function satisfying the WdW equation, some simple analytical extensions of the WdW solutions set constraints into the general features of the probability, time and ensued boundary conditions \cite{Vilenkin94}.
In fact, several hypothesis for circumventing the initial singularity have been suggested such as, for instance, the no-boundary and the tunneling proposals \cite{Hartle83,Linde84,Rubakov84,Vilen84}.

For the HL models considered in this manuscript, an equivalent bounce model is obtained through the extension of the coordinate $a$ (or $x$) from $(0,\infty)$ to $(-\infty,\infty)$, with a quasi-singularity at $a = 0$. 
It means that despite the presence of a potential barrier at $a = 0$, the wave functions from left to right are probabilistically connected.
It can be implemented on the above obtained results by simply suppressing the step-functions $\Theta(x)$ from the integration Eqs~(\ref{eqn27})-(\ref{eqn29}).

\begin{figure}
\vspace{-1 cm}\includegraphics[scale=0.31]{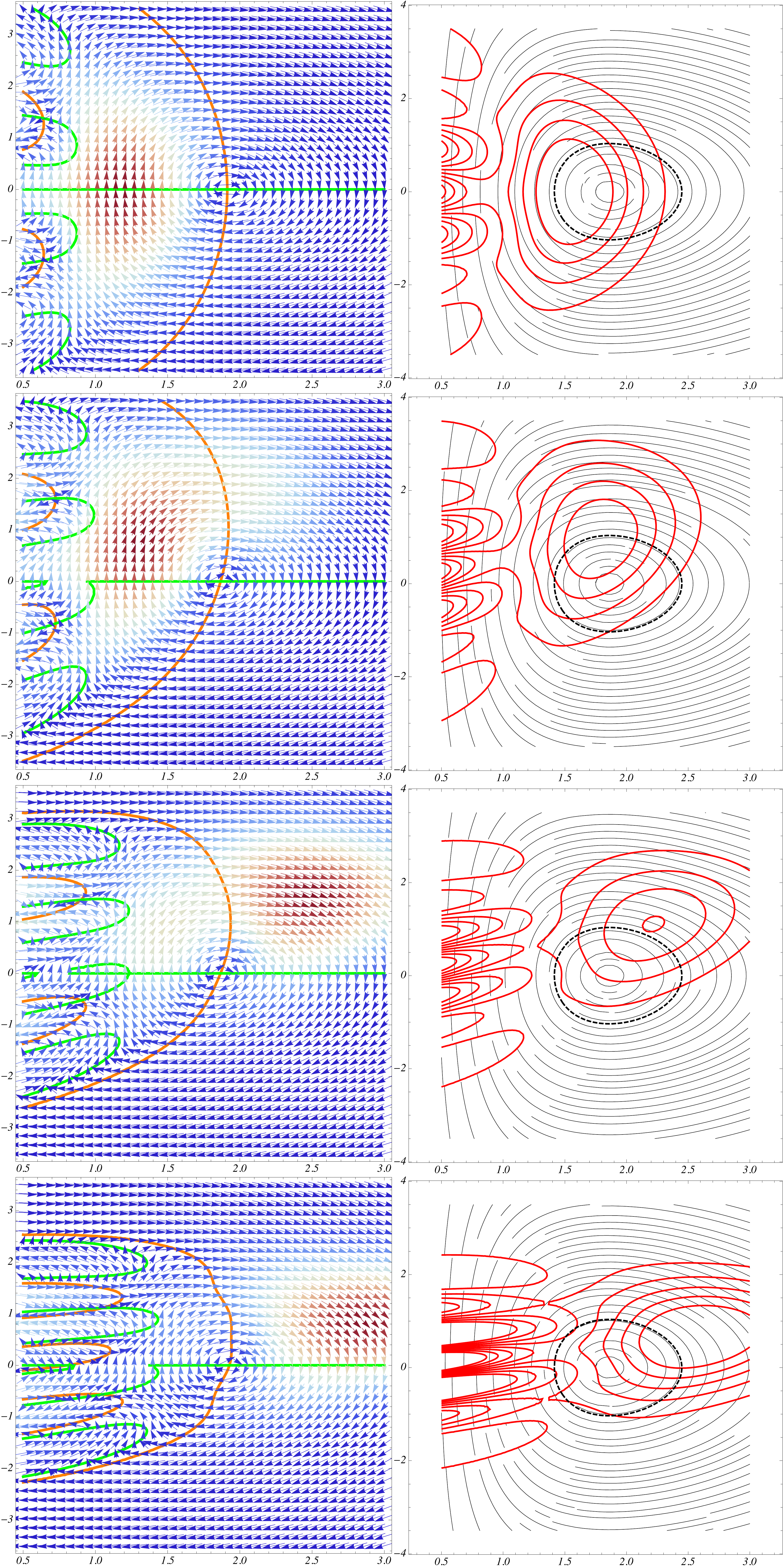}
\renewcommand{\baselinestretch}{.85}
\caption{
(Color online) Bounce HL cosmology with time evolution of the quasi-Gaussian Wigner function, $W^{\alpha}(x,\,p;\, \tau_{})$ in the phase-space $x - p$ plane. Plots are for $\omega \tau_{} = 0$, $\pi/4$, $\pi/2$, and $3\pi/4$, from top to bottom (similar to Fig.~\ref{Figura006}), with $E = 1/2$ and $\alpha = 7/2$. The quantum features exhibited by the contour lines are in correspondence with Fig.~\ref{Figura006}.}
\label{Figura008}
\end{figure}

The corresponding result for the Wigner function should then read 
\begin{eqnarray}
W^{\alpha}_B(x,\,p;\,\tau_{})
&=& \frac{1}{\sqrt{\pi}}\frac{\Gamma(3/2+\alpha)}{\Gamma(1+\alpha)}
(\mu x^2)^{1+\alpha}
\exp\left(-\mu \,x^2\right)\nonumber\\
&&\qquad
\sum_{k=0}^{\frac{1}{2}+\alpha}\frac{x^{-(1+2k)}}{\Gamma(1+k)\,\Gamma(3/2+k+\alpha)}\,
\frac{d^k}{d\mu ^k}\left[\mu ^{-{1}/{2}}\exp\left[-\frac{(p+\tilde{\mu }x)^{2}}{\mu }\right]\right],
\end{eqnarray}
which, as can be seen from Fig.~\ref{Figura008}, it introduces some novel quantum features to its time evolution and the associated currents.
In Fig.~\ref{Figura008} the origin of quantum fluctuations on the right-hand side is due to the probabilistic connection to the left-hand side. 
Even with such quantum fluctuations, for semi-integer values of $\alpha$ one has exactly the same result for the purity of the associated quantum superpostion.
As before, the local fluctuations does not affect the global evolution of the purity for such quantum superpositions.

\section{Perturbative inclusion of the cosmological constant and the age of the Universe}

\begin{figure}
\vspace{-.1 cm}\includegraphics[scale=0.4]{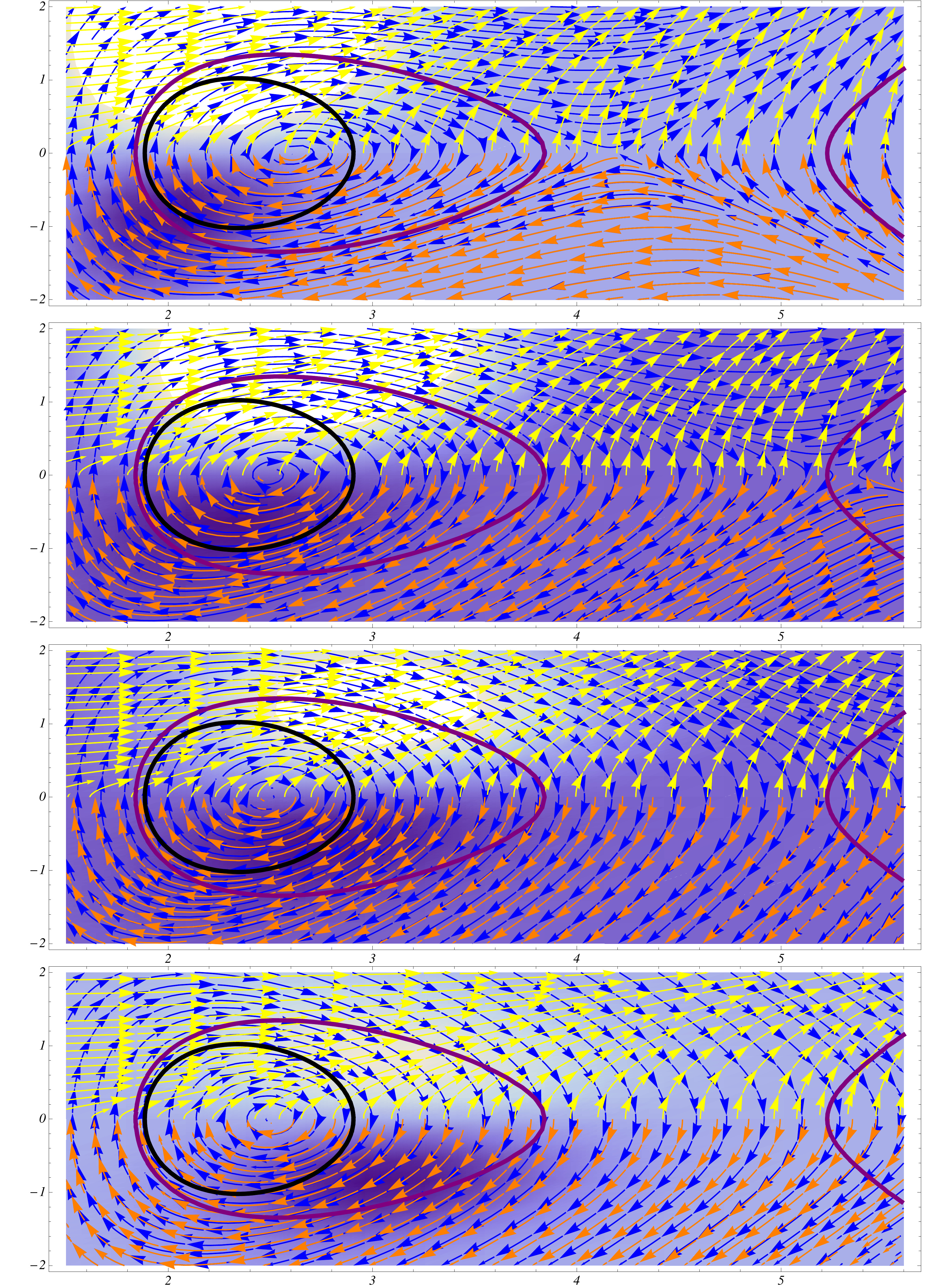}
\renewcommand{\baselinestretch}{.88}
\caption{
(Color online) Complete profile of the Wigner flow (streamlines) for the perturbed HL quantum system in the presence of a cosmological constant.
The Wigner function background color scheme denotes higher values for lighter regions and lower values darker regions.
The amplitude of the Wigner function modulates the (unscaled) arrows.
Streamlines indicate the Wigner flow (normalized to each point) for {\em transmitted} (yellow arrows, $p > 0$) and {\em reflected} (orange arrows, $p < 0$) currents (since they are normalized to each point, the streamline plot corresponds to 
values of ${\bf J}/|{\bf J}|$, therefore only the flow direction is relevant). Results are for $E=2$, $\alpha = 11/2$, and $\ell = 0.022$ (for convenience) and for $\omega\tau_{} = \pi/4$, $\pi/2$, $3\pi/4$ and $\pi$, from top to bottom.}
\label{Figura010}
\end{figure}

Let one now turn back to the expressions for the Wigner currents, Eqs.(\ref{eqn502})-(\ref{eqn505}).
The inclusion of perturbative contributions due to $\ell\,x^4$ leads to the following additional contribution to the Wigner current, 
\footnotesize\begin{eqnarray}
J^{\alpha(\ell)}_p(x,\,p;\,\tau_{})
&=& \frac{\ell}{2}\left(2x^3 - \frac{2\times3\times2}{2\times 3!}x\,\left(\frac{\partial~}{\partial p}\right)^{2}\right)W^{\alpha}(x,\,p;\, \tau_{}) 
\nonumber\\
&=& \frac{\ell}{2}\left(2x^3 \,W^{\alpha}(x,\,p;\, \tau_{})+ \frac{2x}{\pi} 
\int_{-\infty}^{+\infty}dy\,y^2\,\exp(2\,i\,p\,y) \,F^*_{\alpha}(x+y)\,\mathrm{F}_{\alpha}(x-y)\right)
\nonumber\\
&=& \frac{\ell}{2}\left(4x^3\,W^{\alpha}(x,\,p;\, \tau_{}) - \frac{2x}{\pi} 
\int_{-\infty}^{+\infty}dy\,(x^2- y^2)\,\exp(2\,i\,p\,y) \,F^*_{\alpha}(x+y)\,\mathrm{F}_{\alpha}(x-y)\right)
\nonumber\\
&=& 2\ell x^3\left(W^{\alpha}(x,\,p;\, \tau_{}) - \frac{1}{2x^2}\frac{1}{\pi} 
\int_{-\infty}^{+\infty}dy\,(x^2- y^2)\,\exp(2\,i\,p\,y) \,F^*_{\alpha}(x+y)\,\mathrm{F}_{\alpha}(x-y)\right)
\nonumber\\
&=& 2\ell x^3
\left(
W^{\alpha}(x,\,p;\, \tau_{}) - \frac{1}{2 \mu \,x^2}\frac{\alpha+2}{\Gamma(\alpha+1)} W^{\alpha+1}(x,\,p;\, \tau_{})\right).
\label{eqn506}
\end{eqnarray}
\normalsize
Since the background structure of $W^{\alpha}(x,\,p;\, \tau_{})$ does not change, the results due to Eq.~(\ref{eqn506}) can only be considered perturbatively.
Fig.~\ref{Figura010} shows the time evolution of the Wigner flow (background arrows) for the perturbed HL quantum superposition in the presence of a cosmological constant parameterized by $\ell$.
Notice that lighter regions correspond to the probabilistically more relevant components of the Wigner flow bound by the unperturbed Wigner function, $W^{\alpha}(x,\,p;\, \tau_{})$, that moves periodically according to $\mu_{(\beta,\tau_{})}$ and $\tilde{\mu }_{(\beta,\tau_{})}$.
The added perturbation due to the cosmological constant (proportional to minus $x^{4}$) yields conditions for quantum tunneling, which can be observed for times $\sim\omega \tau_{}$ as depicted in Fig.~\ref{Figura010} for $\ell = 0.022$. 
In the pictorial representation shown in Fig.~\ref{Figura010}, the Wigner function approaches the perturbative barrier (cf. Fig.~\ref{Figura001}) depicted by lighter regions which extend over a larger region of the phase-space. This means that part of its probabilistic contribution arises from tunneling (or transmission over the barrier) creating a novel cosmological phase driven by the perturbative current components, $J^{\alpha(\ell)}_p(x,\,p;\,\tau_{})$ (outside the barrier).

To quantitatively understand the above described dynamics \cite{Tunnel}, one can identify the total Wigner flow momentum direction component described by $J^{\alpha(Tot)}_p(x,\,p;\,\tau_{}) = J^{\alpha}_p(x,\,p;\,\tau_{})+ J^{\alpha(\ell)}_p(x,\,p;\,\tau_{})$, with
the contributions for incident (background blue arrows for $p > 0$), transmitted (yellow arrows for $p > 0$) and reflected (orange arrows for $p < 0$) as depicted in Fig.~\ref{Figura010}, such that
\begin{eqnarray}
\mbox{incident flow} &\rightarrow& \Theta(+p)\,J^{\alpha}_p(x,\,p;\,\tau_{}),\nonumber\\
\mbox{transmitted flow} &\rightarrow& \Theta(-p)\,J^{\alpha(\ell)}_p(x,\,p;\,\tau_{}),\nonumber\\
\mbox{reflected flow} &\rightarrow& \Theta(-p)\,(J^{\alpha}_p(x,\,p;\,\tau_{}) + J^{\alpha(\ell)}_p(x,\,p;\,\tau_{})).\nonumber
\label{eqn507}
\end{eqnarray}

By observing that any component $J^{\alpha}_p$ has an even parity with respect to $p \to -p$, and that $\Theta(+p) + \Theta(-p) = 1$, with $\partial_p \Theta(+p) = - \partial_p \Theta(-p)$, one has
\begin{eqnarray}
\int_{-\infty}^{+\infty}\hspace{-.3 cm}dp\,\partial_p J^{\alpha(Tot)}_p &=&
\int_{-\infty}^{+\infty}\hspace{-.3 cm}dp\,\partial_p [\Theta(+p)\,J^{\alpha}_p + \Theta(-p)\,(J^{\alpha}_p + J^{\alpha(\ell)}_p) + \Theta(+p) J^{\alpha(\ell)}_p]\nonumber\\
 &=&
\int_{0}^{+\infty}\hspace{-.3 cm}dp\,\partial_p J^{\alpha}_p - \int_{0}^{+\infty}\hspace{-.3 cm}dp\,\partial_p(J^{\alpha}_p + J^{\alpha(\ell)}_p) - \int_{0}^{+\infty}\hspace{-.3 cm}dp\,\partial_p J^{\alpha(\ell)}_p.
\label{eqn508}
\end{eqnarray}

In addition, from the continuity equation for the momentum component (cf. Eq.~(\ref{eqn53}) in the Appendix II),
one has
\begin{eqnarray}
\frac{d~}{d\tau_{}} \vert G(p;\,\tau_{})\vert^2 &=& \int_{0}^{+\infty}\hspace{-.2 cm}{dx}\,\partial_p J_p(x,\,p;\,\tau),
\label{eqn509}
\end{eqnarray}
which vanishes\footnote{It has been fit to the adequate interval of $x \in (0,\infty)$.} after a symmetric integration over $p$.
By substituting the result from Eq.~(\ref{eqn508}) into the $p$-integrated version of Eq.~(\ref{eqn509}) one has
\begin{eqnarray}
\int_{0}^{+\infty}\hspace{-.2 cm}{dx} \left[\int_{0}^{+\infty}\hspace{-.3 cm}dp\,\partial_p J^{\alpha}_p - \int_{0}^{+\infty}\hspace{-.3 cm}dp\,\partial_p (J^{\alpha}_p + J^{\alpha(\ell)}_p) - \int_{0}^{+\infty}\hspace{-.3 cm}dp\,\partial_p J^{\alpha(\ell)}_p\right] = 0,
\label{eqn510}
\end{eqnarray}
where, in the integrand, the first term is associated to the incident probability which, of course, is shown to be equal to unity, the second term is associated to the reflection probability, $R$, and the last term is associated to the transmission probability, $T$, so that one consistently obtains $R + T=1$.

The quantity $T$ written in terms of
\begin{eqnarray}
T \equiv T(\omega\tau_{}) &=& \int_{0}^{+\infty}\hspace{-.2 cm}{dx} \int_{0}^{+\infty}\hspace{-.3 cm}dp\,\partial_p J^{\alpha(\ell)}_p (x,\,p;\,\tau_{})\nonumber\\
&=& \int_{0}^{+\infty}\hspace{-.2 cm}{dx} ( J^{\alpha(\ell)}_p (x,\,\infty;\,\tau_{}) - J^{\alpha(\ell)}_p (x,\,0;\,\tau_{}))
\nonumber\\
&=& - \int_{0}^{+\infty}\hspace{-.2 cm}{dx} J^{\alpha(\ell)}_p (x,\,0;\,\tau_{})
\label{eqn511}
\end{eqnarray}
defines the {\em transmission (or decaying) rate}, from which one can compute the age of the Universe.

By substituting Eq.~(\ref{eqn506}) into Eq.~(\ref{eqn511}) one obtains
\begin{eqnarray}
T(\omega\tau_{}) &=& 
\int_{0}^{+\infty}\hspace{-.2 cm}{dx} \int_{0}^{+\infty}\hspace{-.3 cm}dp\,\partial_p
 J^{\alpha(\ell)}_p (x,\,p;\,\tau_{})\nonumber\\
&=& - 2\ell\int_{0}^{+\infty}\hspace{-.2 cm}{dx}\,
x^3\left(
W^{\alpha}(x,\,0;\, \tau_{}) - \frac{1}{2 \mu x^2}\frac{\alpha+2}{\Gamma(\alpha+1)} W^{\alpha+1}(x\,0;\, \tau_{})
\right),
\label{eqn512}
\end{eqnarray}
which results into a symbolic expression that be approximated by a constant value given by
\begin{equation}
T(\omega\tau_{}) \simeq 2\ell \times 10^{\frac{3\alpha}{10} -2} = 2g_{_{C}}^{{1}/{2}}\, \frac{g_{_{\Lambda}}}{g_{_{C}}^2} \times 10^{\frac{3\alpha}{10} -2} = \frac{g_{_{C}}^{{1}/{2}}}{18\pi^{2}}\frac{\Lambda}{M_{\mbox{\tiny Pl}}^2} \times 10^{\frac{3\alpha}{10} -2},
\label{eqn513}
\end{equation}
where, in the last step, the quantities have been written in terms of Planck units and
 $$\frac{g_{_{\Lambda}}}{g_{_{C}}^2} = \frac{1}{36\pi^{2}} \frac{\Lambda}{M_{\mbox{\tiny Pl}}^2}.$$
Given that the transmission probability through the barrier is provided by the cosmological constant contribution to the cosmic inventory, $\Omega_{_{\Lambda}}$, then
\begin{equation}
T(\omega\tau_{}) \simeq
\frac{\Omega_{_{\Lambda}}}{\omega\tau_{}} =
\frac{1}{2g_{_{C}}^{{1}/{2}}}\frac{ \Omega_{_{\Lambda}}}{\tau}.
\end{equation}
The contribution, $\Omega_{_{\Lambda}}$, is estimated to be
\begin{equation}
\Omega_{_{\Lambda}} = \frac{g_{_{C}}\,g_{0}}{36\pi^{2}}\times 10^{\frac{3\alpha}{10} -2}\times \tau_{Age},
\label{eqn515}
\end{equation}
where $\Lambda = g_{0} M_{\mbox{\tiny Pl}}^{2}/2$.
Finally, as to recover the current (phenomenological) age of the Universe, $\tau_{Age} \simeq 8\times 10^{60}T_{\mbox{\tiny Pl}}$, for $\Omega_{_{\Lambda}} \simeq 0.7$, from typical values, $g_{_{C}} \simeq 1$ and $g_0\simeq 10^{-123}$, one should have $10^{\frac{3\alpha}{10} -2} \simeq 1.5 \times 10^{61}$, which leads to $\alpha \sim 210$ and is not affected the contribution from {\em radiation} and {\em stiff matter} components since $\Omega_R \sim \sqrt{\Omega_S}$ for large values of $\alpha$.

\section{Conclusions}

The Wigner function and the corresponding Wigner flow for the HL quantum cosmology in the minisuperspace approach have been examined so to visualize the transition from quantum to classical cosmological behaviors in the presence of radiation, curvature and stiff matter components.
In particular, it has been shown that a quantum mechanical parameterization of time as the variable canonically conjugated to the radiation energy density contribution to the HL Hamiltonian is consistent with the classical time evolution.
 
In general, the classical limit for quantum cosmologies described by HL minisuperspace models are difficult to obtain due to the pattern of oscillations of the Wigner functions.
To study the transition from quantum to classical cosmology, the HL model has been described in terms of its Wigner currents.
Our result shows that the averaged quantities obtained from the quasi-Gaussian Wigner function built from the solutions of the WdW equation for the HL quantum mechanical problem coincide with the equivalent classical trajectory of the Universe.
In particular, the classical trajectory matches almost perfectly the maxima of the peak of the Wigner function, with increasingly high accuracy for higher values of the parameter $\alpha$, associated to the stiff matter contribution.

Besides providing an identification with the corresponding classical cosmology, the exact expressions for the Wigner currents show an approximated Liouvillian character for the Wigner flow. 
In addition, a cosmological constant contribution breaks the oscillatory behavior of the Wigner flow due to radiation, curvature and stiff matter contributions and introduces a novel quantum decaying scenario which allows for estimating the age of the Universe.
Moreover, an extension for a kind of bounce model, which extends the space-like coordinate limit from $-\infty$ to $+\infty$, reveals an interesting pattern of quantum interference that produces an identifiable distortion of the Wigner flow.

Finally, it is worth mentioning that the procedure discussed here can be extended to other quantum cosmological scenarios as, for instance, to Kantowski-Sachs \cite{Catarina8,Compean} and modular quantum cosmologies \cite{Schiappa}.
Furthermore, the problem of reaching the classical limit through coarse-graining arguments can be considered by the inclusion of additional friction and diffusion terms in extended versions of our proposal through a deformed quantization formalism.

\vspace{.5 cm}
{\em Acknowledgments} -- The work of AEB is supported by the Brazilian agency FAPESP (grant 17/02294-2). The work of PL is supported by FCT (Funda\c{c}\~ao para a Ci\^encia e a Tecnologia, Portugal) under the grant PD/BD/135005/2017. The work of OB is partially supported by the COST Quantum Structure of Space-Time action MP1405.

\section*{Appendix I -- Normalization and Purity of the Wigner function, $W^{\alpha}(x,\,p;\,\tau_{})$}

From Eq.~(\ref{eqn27}), the normalization condition over $W^{\alpha}(x,\,p;\,\tau_{})$ can be verified through some simple mathematical manipulations.
Firstly, one notices that
\begin{eqnarray}
\int_{0}^{\infty}\hspace{-.3 cm}dx\int_{-\infty}^{+\infty}\hspace{-.3 cm}dp \,W^{\alpha}(x,\,p;\,\tau_{}) &=& \frac{2\,\mu ^{1+\alpha}}{\pi\,\Gamma(1+\alpha)}\,\int_{0}^{\infty}\hspace{-.3 cm}dx \,x ^{2(1+\alpha)}\,\exp\left(-\mu \, x^2\right)\,\int_{-\infty}^{+\infty}\hspace{-.3 cm}dp
\exp\left(2\,i\,x\,p\,s \right)\times\nonumber\\
&& \qquad\quad\int_{-1}^{+1} ds\,(1-s^2)^{\frac{1}{2}+\alpha}\, \exp\left(-\mu \, x^2\,s^2\right) \exp\left(2i\,\tilde{\mu }\, x^2\,s\right),\quad
\label{eqnA01}
\end{eqnarray}
where $y$ has been parameterized as $ y = x\,s$ and $\mu$ and $\tilde{\mu}$ follows from Eqs.~(\ref{eqn23}) and (\ref{tilde}), respectively.
By observing that
\begin{eqnarray}
\int_{-\infty}^{+\infty}\hspace{-.3 cm}dp
\,\exp\left(2\,i\,x\,p\,s \right) &=& 2\pi\,\delta(2\,x\,s) = \frac{\pi}{\vert x\vert}\delta(s),
\label{eqnA02}
\end{eqnarray}
after substitution into Eq.~(\ref{eqnA01}), an integration over the variable $s$ yields
\begin{equation}
\frac{\pi}{\vert x\vert}\int_{-1}^{+1} ds\,\delta(s) (1-s^2)^{\frac{1}{2}+\alpha}\, \exp\left(-\mu \, x^2\,s^2\right) \exp\left(2i\,\tilde{\mu }\, x^2\,s\right) = \frac{\pi}{\vert x\vert}
\label{eqnA01B},
\end{equation}
and then
\begin{eqnarray}
\int_{0}^{\infty}\hspace{-.3 cm}dx\int_{-\infty}^{+\infty}\hspace{-.3 cm}dp \,W^{\alpha}(x,\,p;\,\tau_{}) &=& \frac{2\,\mu ^{1+\alpha}}{\Gamma(1+\alpha)}\,\int_{0}^{\infty}\hspace{-.3 cm}dx \,x ^{(1+2\alpha)}\,\exp\left(-\mu \, x^2\right) = 1\label{eqnA03}.
\end{eqnarray}

By following a similar strategy, given the purity (cf. Eq.~(\ref{pureza})),
\begin{equation}
\mathcal{P} = 2\pi\int_{0}^{\infty}\hspace{-.3 cm}dx\int_{-\infty}^{+\infty}\hspace{-.3 cm}dp \,\left(W^{\alpha}(x,\,p;\,\tau_{})\right)^{2},
\end{equation}
one notices that
\begin{eqnarray}
\lefteqn{\int_{0}^{\infty}\hspace{-.3 cm}dx\int_{-\infty}^{+\infty}\hspace{-.3 cm}dp \,\left(W^{\alpha}(x,\,p;\,\tau_{})\right)^{2} =\frac{4\,\mu ^{2(1+\alpha)}}{\pi^2\,\Gamma^2(1+\alpha)}\times}\nonumber\\
&& \,\int_{0}^{\infty}\hspace{-.3 cm}dx \,x^{4(1+\alpha)}\,\exp\left(-2\,\mu \, x^2\right)
\,\int_{-\infty}^{+\infty}\hspace{-.3 cm}dp
\,\exp\left(2\,i\,x\,p \,(s+r) \right)\times
\nonumber\\
&& \qquad\int_{-1}^{+1} ds\int_{-1}^{+1} dr\,[(1-r^2)(1-s^2)]^{\frac{1}{2}+\alpha}\, \exp\left(-\mu \, x^2\,(r^2+s^2)\right) \exp\left(2i\,\tilde{\mu }\, x^2\,(s+r)\right). \quad\,
\label{eqnA04}
\end{eqnarray}
After substituting
\begin{eqnarray}
\int_{-\infty}^{+\infty}\hspace{-.3 cm}dp
\,\exp\left(2\,i\,x\,p \,(s+r) \right) &=& 2\pi\,\delta(2\,x\,(s+r)) = \frac{\pi}{\vert x\vert}\delta(s+r)
\label{eqnA05}
\end{eqnarray}
into Eq.~(\ref{eqnA04}), an integration over the variable $r$ gives
\begin{eqnarray}
\lefteqn{\int_{0}^{\infty}\hspace{-.3 cm}dx\int_{-\infty}^{+\infty}\hspace{-.3 cm}dp \,\left(W^{\alpha}(x,\,p;\,\tau_{})\right)^{2} = \frac{4\,\mu ^{2(1+\alpha)}}{\pi^2\,\Gamma^2(1+\alpha)}\times}\nonumber\\ &&\qquad\qquad\int_{0}^{\infty}\hspace{-.3 cm}dx \,x^{(3+4\alpha)}\,\exp\left(-2\,\mu \, x^2\right)\,\int_{-1}^{+1} ds\,(1-s^2)^{1+2\alpha}\, \exp\left(-2\,\mu \, x^2\,s^2\right).
\label{eqnA05}
\end{eqnarray}
By evaluating the integral over $x$, one has
\begin{eqnarray}
\int_{0}^{\infty}\hspace{-.3 cm}dx \,x^{(3+4\alpha)}\,\exp\left(-2\,\mu \, x^2\,(1+s^2)\right) = \frac{1}{2^{3+2\alpha}\mu ^{2(1+\alpha)}}
\frac{\Gamma(2(1+\alpha))}{(1+s^2)^{2(1+\alpha)}},
\label{eqnA06}
\end{eqnarray}
which can be substituted into Eq~(\ref{eqnA05}) as to give
\begin{eqnarray}
\int_{0}^{\infty}\hspace{-.3 cm}dx\int_{-\infty}^{+\infty}\hspace{-.3 cm}dp \,\left(W^{\alpha}(x,\,p;\,\tau_{})\right)^{2}
 &=&
\frac{1}{2^{1+2\alpha}\pi}
\frac{\Gamma(2(1+\alpha))}{\Gamma^2(1+\alpha)} \int_{-1}^{+1} ds\,\frac{(1-s^2)^{1+2\alpha}}{(1+s^2)^{2(1+\alpha)}}\nonumber\\
 &=&
\frac{1}{2^{2+2\alpha}\pi}
\frac{\Gamma(2(1+\alpha))}{\Gamma^2(1+\alpha)} \frac{\sqrt{\pi} \Gamma(1+\alpha)}{2\Gamma(3/2+\alpha)}
\nonumber\\
 &=&\frac{1}{2\pi},
\label{eqnA07}
\end{eqnarray}
that confirms that $\mathcal{P}=1$ for $W^{\alpha}(x,\,p;\,\tau_{})$ defined by Eq.~(\ref{eqn27}).

\section*{Appendix II -- Wigner flow and phase-space quantum information quantifiers} 

Interesting quantum aspects of a physical system can be revealed by the time evolution of the Wigner function \cite{Steuernagel3,Ferraro11}, $W(\xi_x,\,\xi_p;\,t)$, when it is cast in the form of a vector flux $\mathbf{J}(\xi_x,\,\xi_p;\,t)$ \cite{Donoso12,Domcke,Waldron17}.
This flow drives the quasi-probability density in the phase-space as well as it reproduces the dynamics of a quantum system. Written in the form of a continuity equation \cite{Case,Ballentine,Steuernagel3}
\begin{equation}
\frac{\partial W}{\partial t} + \mbox{\boldmath $\nabla$}_{\xi}\cdot \mathbf{J}=\frac{\partial W} {\partial t} + \frac{\partial J_x}{\partial \xi_x}+\frac{\partial J_p}{\partial \xi_p} =
 0,
\label{eqn51}
\end{equation}
through the $\xi_x - \xi_p$ decomposition, $\mathbf{J} = J_x\,\hat{\xi}_x + J_p\,\hat{\xi}_p$, with
\begin{eqnarray}
J_x(\xi_x,\,\xi_p;\,t)&=& \frac{\xi_p}{m}\,W(\xi_x,\,\xi_p;\,t),\\
\label{eqn500BB}
J_p(\xi_x,\,\xi_p;\,t)&=& -\sum_{k=0}^{\infty} \left(\frac{i\,\hbar}{2}\right)^{2k}\frac{1}{(2k+1)!} \, \left(\frac{\partial~}{\partial \xi_x}\right)^{2k+1}\hspace{-.5cm}V(\xi_x)\,\left(\frac{\partial~}{\partial \xi_p}\right)^{2k}\hspace{-.3cm}W(\xi_x,\,\xi_p;\,t),
\label{eqn500}
\end{eqnarray}
the probability density, $\vert \mathrm{F}(\xi_x;\,t)\vert^2$, and the momentum distribution $\vert G(\xi_p;\,t)\vert^2$,
are related one to each other by
\begin{equation}
G(\xi_p;\,t) =(2\pi\hbar)^{-1/2} \int_{-\infty}^{+\infty}\hspace{-.2 cm}{d\xi_x}\,\mathrm{F}(\xi_x;\,t)\,\exp(i\,\xi_p\,\xi_x/\hbar),
\label{eqn52}
\end{equation}
and have a time evolution given by
\begin{equation}
\frac{d}{dt} \vert \mathrm{F}(\xi_x;\,t)\vert^2 = \int_{-\infty}^{+\infty}\hspace{-.2 cm}{d\xi_p}\,\partial_{\xi_x} J_x(\xi_x,\,\xi_p;\,t) = \partial_{\xi_x} j_x(\xi_x;\,t),
\label{eqn53A}
\end{equation}
\begin{equation}
\frac{d}{dt} \vert G(\xi_p;\,t)\vert^2 = \int_{-\infty}^{+\infty}\hspace{-.2 cm}{d\xi_x}\,\partial_{\xi_p} J_p(\xi_x,\,\xi_p;\,t) = \partial_{\xi_p} j_p(\xi_p;\,t).
\label{eqn53}
\end{equation}

Of course, the first of the above equations has a quantum analog in the coordinate representation, that is 
$$\int_{-\infty}^{+\infty}\hspace{-.2 cm}{dp}\,J_x(\xi_x,\,\xi_p;\,t) = j_x(\xi_x;\,t).$$

For the purpose of comparing classical and quantum dynamics, one firstly identifies the classical Hamiltonian phase-space velocity, 
$\dot{\mbox{\boldmath $\xi$}} = \mathbf{v}_{\xi} = (v_x,\,v_p)$, associated to a
coordinate vector $\mbox{\boldmath $\xi$} = (\xi_x,\,\xi_p)$, so to have the classical flow field given by $\mathbf{J} = \mathbf{v}_{\xi}\,W$, with $v_x = \dot{\xi}_x = \xi_p/m$ and $v_p = \dot{\xi}_p = -\partial V/\partial \xi_x$.
The analogy with fluid dynamics is indeed much more intuitive in the classical regime for which Eq.~\eqref{eqn500} reduces to the Liouville equation.
In particular, the classical velocity for time independent conservative systems is divergence free, i.e. $\mbox{\boldmath $\nabla$}_{\xi} \cdot \mathbf{v}_{\xi} = 0$.
In this case, to compute time variation of the integrated probability for a volume of fluid bound by an enclosing path, $\mathcal{C}$, which moves with $\mathbf{v}_{\xi(\mathcal{C})} = (\xi_p/m,\, -\partial V/\partial \xi_x)$, Eq.~\eqref{eqn51} is cast in the form of
\begin{eqnarray}
\frac{\partial W}{\partial t} + \mbox{\boldmath $\nabla$}_{\xi}\cdot(\mathbf{v}_{\xi(\mathcal{C})}\,W )&=& 0, \qquad (classical).\label{classical}\end{eqnarray}
By observing that the convective derivative operator \cite{Steuernagel3} corresponds to 
\begin{equation}
\frac{D~}{Dt} \equiv \frac{\partial~}{\partial t} + \mathbf{v}_{\xi} \cdot \mbox{\boldmath $\nabla$}_{\xi},
\label{eqn57}
\end{equation}
one can rewrite the continuity equation, Eq.~\eqref{classical}, as
\begin{equation}
\frac{D W}{Dt} = - W\, \mbox{\boldmath $\nabla$}_{\xi} \cdot \mathbf{v}_{\xi},
\label{eqn57B}
\end{equation}
which corresponds to a conservation law whenever $\frac{D W}{Dt} = 0$.
It establishes that the fluid-analog is Liouvillian and incompressible.

Otherwise, for the quantum case described by Eq.~\eqref{eqn51}, $\mathbf{J}$ may be associated to $\mathbf{u}\,W$, and a typical non-Liouvillian \cite{Liouvillian} flow can be recognized through the divergence pattern of $\mathbf{u}$, $\mbox{\boldmath $\nabla$}_{\xi} \cdot \mathbf{u} \neq 0$.
The Wigner phase-velocity, $\mathbf{u}$, the quantum analog of $\mathbf{v}_{\xi(\mathcal{C})}$, exhibits a subtle unbound divergent behavior expressed by
\begin{equation}
\mbox{\boldmath $\nabla$}_{\xi} \cdot \mathbf{u} = \frac{W\, \mbox{\boldmath $\nabla$}_{\xi}\cdot \mathbf{J} - \mathbf{J}\cdot\mbox{\boldmath $\nabla$}_{\xi}W}{W^2},
\label{eqn59}
\end{equation}
where $\mbox{\boldmath $\nabla$}_{\xi}\cdot\mathbf{J} = W\mbox{\boldmath $\nabla$}_{\xi}\cdot\mathbf{u}+ \mathbf{u}\cdot \mbox{\boldmath $\nabla$}_{\xi}W$.
The condition that sets $\mbox{\boldmath $\nabla$}_{\xi} \cdot \mathbf{u} \neq 0$ is quite helpful in identifying an approximated Liouvillian dynamics in the phase-space. 

Besides producing a picture of the local quantum distortions, the operator $\mbox{\boldmath $\nabla$}_{\xi} \cdot \mathbf{u}$ has also a closed relation with the rate of change of the purity \cite{Entro02} as given by
 \cite{MeuPaper}
\begin{equation}
\frac{1}{2\pi}\frac{D\mathcal{P}}{Dt} + \langle W\,\mbox{\boldmath $\nabla$}_{\xi}\cdot\mathbf{u} \rangle =0.
\label{eqn62BB}
\end{equation}

The rate of change of $\mathcal{P}$ is driven by quantum distortions over the background Liouvillian flow.
From Eq.~(\ref{eqn500}), it is possible to demonstrate that the purity is only locally affected since, once integrated, one has
\begin{eqnarray}
\frac{D\mathcal{P}}{Dt} \propto \int_{-\infty}^{+\infty}\hspace{-.3 cm}d\xi_p\, W\,\left(\frac{\partial~}{\partial\xi_ p}\right)^{2k+1}\hspace{-.5 cm} W(\xi_x,\,\xi_p;\,t) = 0,
\label{eqn65}
\end{eqnarray}
i.e. the purity is a constant of the motion if the integration volume is extended over all the phase-space, or even over a symmetric interval in the momentum direction, for the cases where $W$ is symmetric in $\xi_p$.

In what concerns the quantum cosmological solutions discussed in the body of the paper, a quantifier of quantum fluctuations is given by the averaged value of $\Delta J_p(\xi_{x_{_{\mathcal{C}}}}\bb{t},\,\xi_{p_{_{\mathcal{C}}}}\bb{t};t)$ (corresponding to $\Delta J^{\alpha}_p(x,\,p;\,\tau_{})$ for the cosmological canonical variables) over all the phase-space volume.
For periodic motions defined by classical trajectories parametrized by $\mathcal{C}$ \cite{MeuPaper}, one can assign to the classical trajectory the role of a two-dimensional boundary contour for the Wigner flow.
In this case, the integration of the dependent expression for $\partial W/\partial t$ over a volume $V_{\mathcal{C}}$ enclosed by $\mathcal{C}$, results into
\begin{equation}
\int_{V_\mathcal{C}}dV\, \frac{\partial W}{\partial t} =
\int_{V_\mathcal{C}}dV\, \left(\frac{DW}{Dt} - \mathbf{v}_{\xi(\mathcal{C})} \cdot \mbox{\boldmath $\nabla$}_{\xi} W\right) =
\frac{D~}{Dt}\int_{V_\mathcal{C}}dV \,W - \int_{_{\mathcal{C}}}dV \,\mbox{\boldmath $\nabla$}_{\xi}\cdot (\mathbf{v}_{\xi(\mathcal{C})}W), 
\label{eqn51BB}
\end{equation}
which, upon substitution into an integrated version of Eq.~(\ref{eqn51}), leads to
\begin{equation}
\frac{D~}{Dt}\int_{V_\mathcal{C}}dV \,W = \int_{V_\mathcal{C}}dV \,\left(\mbox{\boldmath $\nabla$}_{\xi}\cdot (\mathbf{v}_{\xi(\mathcal{C})}W) - \mbox{\boldmath $\nabla$}_{\xi}\cdot \mathbf{J}\right),
\label{eqn51CC}
\end{equation}
which vanishes in the classical limit, i.e. when $\mathbf{u}\sim\mathbf{v}_{\xi(\mathcal{C})}$.

To identify the quantum corrections in terms of $\Delta \mathbf{J} = \mathbf{J} - \mathbf{v}_{\xi(\mathcal{C})}W$, one computes the variation of the integrated probability flux enclosed by the classical surface, $\mathcal{C}$, in terms of an integral over a path given by
\begin{equation}
\frac{D~}{Dt}\int_{V_\mathcal{C}}dV \,W = \frac{D}{Dt}\mbox{Prob}_{(\mathcal{C})} = -\int_{V_\mathcal{C}}dV\, \mbox{\boldmath $\nabla$}_{\xi}\cdot \Delta \mathbf{J} = -\oint_{_{\mathcal{C}}}d\ell\,\Delta \mathbf{J}\cdot \mathbf{n},
\label{eqn51DD}
\end{equation}
where the unitary vector, $\mathbf{n}= (-\dot{\xi}_p{_{_{\mathcal{C}}}}, \dot{\xi}_{x_{_{\mathcal{C}}}}) \vert\mathbf{v}_{\xi(\mathcal{C})}\vert^{-1}$, is orthogonal to $\mathbf{v}_{\xi(\mathcal{C})}$.
By following the parameterization of the line element, $d\ell \equiv \vert\mathbf{v}_{\xi(\mathcal{C})}\vert dt$, one has
\begin{equation}
\frac{D}{Dt}\mbox{Prob}_{(\mathcal{C})}
\bigg{\vert}_{t = T} = -\oint_{_{\mathcal{C}}}d\ell\, \mathbf{J}\cdot \mathbf{n} = -
\int_{0}^{T}dt\, \Delta J_p(\xi_{x_{_{\mathcal{C}}}}\bb{t},\,\xi_{p_{_{\mathcal{C}}}}\bb{t};t)\,\,\dot{\xi}_{x_{_{\mathcal{C}}}}\bb{t},
\label{eqn51EE}
\end{equation}
where $\xi_{x_{_{\mathcal{C}}}}\bb{t}$ and $\xi_{p_{_{\mathcal{C}}}}\bb{t}$ are typical classical solutions, $T$ is the period of the classical motion, and $\Delta J_p(\xi_x,\,\xi_p;\,t)$ is given by Eq.~\eqref{eqn500} for $k\geq 1$.

\end{document}